\documentclass[journal=jctcce,manuscript=article]{achemso}

\usepackage{natbib}
\usepackage[utf8]{inputenc}
\usepackage[english]{babel}
\usepackage[T1]{fontenc}
\usepackage[bookmarks=false]{hyperref}
\usepackage{amsfonts, amsmath, amssymb, mathrsfs, mathtools, amsthm, bbm, bm}
\usepackage{caption}
\usepackage{subcaption}
\usepackage{braket}
\usepackage{float}
\usepackage[version=4]{mhchem}
\usepackage{tikz}
\usepackage{quantikz}
\usepackage{booktabs, tabularx}
\mciteErrorOnUnknownfalse

\DeclareMathOperator{\Tr}{Tr}
\newcommand{\I}[2][I]{\ensuremath{\mathcal{#1}_\mathrm{#2}}}
\newcommand{\fa}{\;\forall\,}
\newcolumntype{Y}{>{\raggedleft\arraybackslash}X} 

\author{Tim Weaving}
\email{timothy.weaving.20@ucl.ac.uk}
\affiliation{Centre for Computational Science, Department of Chemistry, University College London, WC1H 0AJ, United Kingdom}

\author{Alexis Ralli}
\email{alexis.ralli.18@ucl.ac.uk}
\affiliation{Centre for Computational Science, Department of Chemistry, University College London, WC1H 0AJ, United Kingdom}

\author{William M. Kirby}
\email{william.kirby@tufts.edu}
\affiliation{Department of Physics and Astronomy, Tufts University, Medford, MA 02155, USA}

\author{Andrew Tranter}
\email{main@atranter.net}
\affiliation{Cambridge Quantum Computing, 9a Bridge Street Cambridge, CB2 1UB, United Kingdom}

\author{Peter J. Love}
\email{peter.love@tufts.edu}
\affiliation{Department of Physics and Astronomy, Tufts University, Medford, MA 02155, USA}
\altaffiliation{Computational Science Initiative, Brookhaven National Laboratory, Upton, NY 11973, USA}

\author{Peter V. Coveney}
\email{p.v.coveney@ucl.ac.uk}
\affiliation{Centre for Computational Science, Department of Chemistry, University College London, WC1H 0AJ, United Kingdom}
\altaffiliation{UCL Centre for Advanced Research Computing, Gower Street, London WC1E 6BT, United Kingdom}
\altaffiliation{Informatics Institute, University of Amsterdam, Amsterdam, 1098 XH, Netherlands}

\title{A Stabilizer Framework for Contextual Subspace VQE and the Noncontextual Projection Ansatz}

\begin{document}

\begin{abstract}
Quantum chemistry is a promising application for noisy intermediate-scale quantum (NISQ) devices. However, quantum computers have thus far not succeeded in providing solutions to problems of real scientific significance, with algorithmic advances being necessary to fully utilise even the modest NISQ machines available today. We discuss a method of ground state energy estimation predicated on a partitioning of the molecular Hamiltonian into two parts: one that is \textit{noncontextual} and can be solved classically, supplemented by a \textit{contextual} component that yields quantum corrections obtained via a Variational Quantum Eigensolver (VQE) routine. This approach has been termed \textit{Contextual Subspace VQE} (CS-VQE); however, there are obstacles to overcome before it can be deployed on NISQ devices. The problem we address here is that of the ansatz, a parametrized quantum state over which we optimize during VQE; it is not initially clear how a splitting of the Hamiltonian should be reflected in the CS-VQE ans\"atze. We propose a `noncontextual projection' approach that is illuminated by a reformulation of CS-VQE in the stabilizer formalism. This defines an ansatz restriction from the full electronic structure problem to the contextual subspace and facilitates an implementation of CS-VQE that may be deployed on NISQ devices. We validate the noncontextual projection ansatz using a quantum simulator and demonstrate chemically accurate ground state energy calculations for a suite of small molecules at a significant reduction in the required qubit count and circuit depth. \end{abstract}

\section{Introduction}

Quantum computers promise to yield solutions to complex problems that have previously been unattainable by classical means, yet experimental demonstration remains challenging. To date, the largest molecules simulated on noisy intermediate-scale quantum (NISQ) hardware are \ce{H12} -- albeit only a Hartree-Fock calculation -- conducted by Google using just 12 of the 53 qubits available on their superconducting quantum processor \textit{Sycamore} \cite{arute2020hartree}, and \ce{H2O}, performed independently by IonQ using 3 qubits of an unspecified proprietary trapped ion device \cite{nam2020ground} and IBM, using 5 of the 27 qubits on the now-decommissioned \textit{ibmq\_dublin} superconducting device \cite{eddins2021doubling}.

Due to the limitations of shallow circuit depth and short coherence times that characterise the NISQ era, we are not able to harness the full state-space afforded to these machines. To circumvent the above issues, we turn to the class of variational quantum algorithms, of which the Variational Quantum Eigensolver (VQE) \cite{peruzzo2014variational} is most widely studied. In contrast with eigenvalue-finding algorithms requiring fault-tolerant machines such as Quantum Phase Estimation (QPE) \cite{kitaev1995quantum} -- which necessitates state evolution over an extended period of coherence -- VQE executes a large ensemble of comparatively shallow parametrized circuits to estimate energy expectation values, informing a classical optimizer that updates the parameter settings before reinitialization of the quantum circuit. Its success is predicated on the variational principle, meaning the ground state energy of the system bounds expectation values from below \cite{griffiths2018introduction}.

However, VQE is not without its challenges. First of all, the parametrized quantum state mentioned above -- known as an \textit{ansatz} -- needs to be constructed carefully; it must be sufficiently expressible so the subspace of quantum states it spans contains the true ground state. On the other hand, if the ansatz is too expressible, we run into the problem of barren plateaus \cite{mcclean2018barren} where we observe vanishing gradients. This is more often a symptom of `hardware efficient' ans\"atze \cite{arute2020hartree, wecker2015progress, kandala2017hardware, rattew2019domain, wiersema2020exploring, harrigan2021quantum}, which aim to access the largest possible region of Hilbert space for the fewest number of native quantum gates.

To avoid barren plateaus, one must take into account some of the underlying problem structure to define ansatz circuits whose image is confined to a smaller, but more targetted, region of Hilbert space. Within this category are `chemically inspired' ans\"atze that represent sequences of electronic excitation operators in circuit; unitary coupled cluster (UCC) \cite{taube2006new, romero2018strategies} is widely acknowledged as the gold standard for electronic structure simulations, albeit computationally very expensive in practice.

More recently, we have seen the development of hybrid ans\"atze that bridge the gap between hardware efficiency and chemical motivation. For example, Gard et al. \cite{gard2020efficient} designed a compact circuit designed to conserve molecule symmetries such as particle number and spin, while Adaptive Derivative-Assembled Pseudo-Trotter (ADAPT) VQE \cite{grimsley2019adaptive, tang2021qubit, shkolnikov2021avoiding, fedorov2021unitary} describes a more complete approach to scalable quantum chemistry simulations by defining selection criteria of ansatz terms from a pool of excitation operators.

Secondly, the energy estimation procedure in VQE invokes the measurement problem; in order to mitigate statistical error, many prepare-and-measure cycles are necessary to achieve sufficient precision in the estimate. The advances made in recent years towards measurement reduction techniques are expansive \cite{babbush2018low, wang2019accelerated, verteletskyi2020measurement, jena2019pauli, gokhale2019minimizing, yen2020measuring, huggins2021efficient, gokhale2020n, torlai2020precise, crawford2021efficient} and range from classical pre/post-processing of the measurement information such as in classical shadow tomography \cite{huang2020predicting, hadfield2020measurements} to Hamiltonian term-grouping schemes and reductions in the number of Hamiltonian terms at a cost of coherent resource, such as in unitary partitioning \cite{izmaylov2019unitary, zhao2020measurement, bonet2020nearly, ralli2021implementation, ralli2022unitary}. Combined with techniques of error mitigation \cite{temme2017error, li2017efficient, endo2018practical, kandala2019error, giurgica2020digital, he2020zero, endo2021hybrid, huggins2021virtual}, one can optimize VQE with the objective of maximal NISQ resource utilization.

%We are at a critical point in the development of quantum computation, where practical methods of quantum computing must be at the forefront to drive scientific innovations outside the realm of theoretical curiosities. Quantum biology and chemistry are the main areas of research due to benefit from the advances in quantum simulation techniques, with the elucidation of previously intractable biological and chemical mechanisms no longer being out of the question.

In this work we are concerned with Contextual Subspace VQE (CS-VQE) \cite{kirby2021contextual}, which describes a method of partitioning the molecular Hamiltonian into disjoint parts so that an electronic structure problem may be simulated to some degree on the available quantum device, even when the dimension of the full problem is too great to be encoded on the number of qubits available. This is supplemented by some classical overhead, but this often permits one to achieve chemical accuracy ($1.6\mathrm{mHa} \approx 4\mathrm{kJ/mol}$) at a saving of qubits, as indicated by Kirby et al. \cite{kirby2021contextual}.

There has since been further research into the use of classical estimates of the electronic structure problem to reduce the resource requirements on quantum hardware. In particular, Classically-Boosted VQE (CB-VQE) \cite{radin2021classically} identifies classically tractable states and excludes them from the quantum simulation, alleviating some measurement and fidelity requirements of the VQE routine.

CS-VQE bears a resemblance to the qubit reduction technique of qubit tapering \cite{bravyi2017tapering, setia2020reducing}, which exploits $\mathbb{Z}_2$ symmetries of the Hamiltonian; the differences and similarities are highlighted herein and by Kirby et al. \cite{kirby2021contextual}. However, there are still a number of problems to address before CS-VQE may be successfully deployed on real quantum hardware, most notably with regard to the ansatz, which is the principal focus of this work.

\section{Preliminaries}\label{prelim}

The notation used throughout shall be to write operators in standard capital font ($A, B, C, \dots$), with the exception of single-qubit Pauli operators being written in the form
\begin{equation}
    \sigma_p \coloneqq \begin{pmatrix} \delta_{p,0} + \delta_{p,3} & \delta_{p,1} - i \delta_{p,2} \\ \delta_{p,1} + i \delta_{p,2} & \delta_{p,0} - \delta_{p,3} \end{pmatrix}
\end{equation}
for $p \in \{0,1,2,3\}$. Sets are denoted by calligraphic letters ($\mathcal{A}, \mathcal{B}, \mathcal{C}, \dots$) and vector spaces by script typeface ($\mathscr{A}, \mathscr{B}, \mathscr{C}, \dots$).
The state space of $N$ qubits may be identified with the $2^N$ dimensional Hilbert space $\mathscr{H} = (\mathbb{C}^2)^{\otimes N}$, with the space of (bounded) linear operators acting upon $\mathscr{H}$ denoted $\mathscr{B}(\mathscr{H})$. 

We introduce the \textit{Pauli group} $\mathcal{P}_N \subset \mathscr{B}(\mathscr{H})$ consisting of operators $\bm{\sigma}=\bigotimes_{i = 0}^{N-1} \sigma_{p_i}$ for $p_i \in \{0,1,2,3\}$, up to multiplication by $\pm 1, \pm i$. Note the distinction between the bold font $\bm{\sigma}$ denoting tensor products and $\sigma_p$ a single-qubit Pauli operator; we will sometimes write $\sigma_p^{(i)}$ to index explicitly the qubit position $i \in \mathbb{Z}_N$ on which it acts. We shall also make use of the \textit{commutator} $[A, B] \coloneqq AB - BA$ and \textit{anticommutator} $\{A, B\} \coloneqq AB + BA$, defined for operators $A,B \in \mathscr{B}(\mathscr{H})$, which are zero when $A$ and $B$ commute/anticommute, respectively.

An $N$-qubit Hamiltonian can be written in the form
\begin{equation}\label{hamiltonian}
    H_{\mathcal{T}} = \sum_{\bm{\sigma}\in\mathcal{T}} h_{\bm{\sigma}} \bm{\sigma},\;\, h_{\bm{\sigma}} \in \mathbb{R},
\end{equation}
for a set of Pauli operators $\mathcal{T} \subset \mathcal{P}_N$ -- specifying real coefficients ensures $H_{\mathcal{T}}$ is \textit{Hermitian}. The objective of quantum chemistry simulations is to estimate the ground state energy
\begin{equation}
    \epsilon_0 \coloneqq \min_{\ket{\psi} \in \mathscr{H}} \braket{H_\mathcal{T}}_\psi
\end{equation}
where $\braket{H_\mathcal{T}}_\psi \coloneqq \bra{\psi} H_\mathcal{T} \ket{\psi}$ is the expectation value of $H_\mathcal{T}$ with respect to some quantum state $\ket{\psi} \in \mathscr{H}$. Many physical properties of the target system are determined by the ground state, motivating this goal.

The Variational Quantum Eigensolver (VQE) quantum-classical hybrid algorithm \cite{peruzzo2014variational} is the most widely studied means of achieving this on NISQ hardware. VQE requires a parametrized ansatz state $\ket{\psi_\mathrm{anz}(\bm{\theta})} \in \mathscr{H}$ whose parameters $\bm{\theta}$ are manipulated within a classical optimization scheme that aims to minimize the energy expectation value
\begin{equation}\label{expval}
    E(\bm{\theta}) \coloneqq \braket{H_\mathcal{T}}_{\psi_\mathrm{anz}(\bm{\theta})},
\end{equation}
evaluated via many prepare-and-measure cycles. The choice of ansatz restricts us to a subspace of quantum states and therefore must be carefully designed to be sufficiently expressible so as to capture the true ground state of the system.

A common form of ansatz state -- particularly in relation to the electronic structure problem -- is
\begin{equation}\label{vanilla_anz}
    \ket{\psi_\mathrm{anz}(\bm{\theta})} = e^{iA(\bm{\theta})} \ket{\psi_\mathrm{ref}},    
\end{equation}
where $\ket{\psi_\mathrm{ref}} \in \mathscr{H}$ is some fixed reference state in which the quantum circuit is initialized and $A(\bm{\theta})=\sum_{\bm{\sigma \in \mathcal{A}}} \theta_{\bm{\sigma}} \bm{\sigma}$ for parameters $\theta_{\bm{\sigma}} \in \mathbb{R}$ and Pauli operators $\bm{\sigma} \in \mathcal{A} \subset \mathcal{P}_N$; the unitary $e^{iA(\bm{\theta})}$ effects excitations above the reference state. Such ans\"atze as unitary coupled cluster (UCC) \cite{taube2006new, romero2018strategies} may by expressed by our choice of $A$ (taking as reference the Hartree-Fock state), in addition to any others based on the theory of excitation operators such as ADAPT-VQE \cite{grimsley2019adaptive, tang2021qubit, shkolnikov2021avoiding, fedorov2021unitary}. The quantum advantage in VQE stems from the ability to prepare classically intractable states from our parametrized ansatz circuits.

\section{Projections onto stabilizer subspaces}\label{stabilizers}

%An idea that is commonplace within qubit reduction techniques is that of projecting the Hamiltonian onto a stabilizer subspace spanning some subset of qubits. 
Given an operator $\bm{\sigma} \in \mathcal{P}_N$, the space of quantum states $\ket{\psi} \in \mathscr{H}$ that it stabilizes are those satisfying $\bm{\sigma}\ket{\psi}=\ket{\psi}$, the $+1$-eigenspace of $\bm{\sigma}$. Extending this notion to an abelian subgroup of Pauli operators $\mathcal{Q} \subset \mathcal{P}_N$, there is an induced vector space $\mathscr{V}_\mathcal{Q}$ of states stabilized by the elements of $\mathcal{Q}$.

A particularly useful definition is that of a Hamiltonian \textit{symmetry}, taken here to mean a set $\mathcal{S} \subset \mathcal{P}_N$ of Pauli operators such that
\begin{equation}
    [\bm{\sigma}, \bm{\sigma}^\prime]=0 \;\fa \bm{\sigma}\in\mathcal{S}, \bm{\sigma}^\prime\in\mathcal{T}.    
\end{equation}
In other words, a symmetry of $H_\mathcal{T}$ is any set of Pauli operators that commute universally among $\mathcal{T}$, which we may extend to an abelian group $\overline{\mathcal{S}} \coloneqq \braket{\mathcal{S}}$ generated by $\mathcal{S}$ under operator multiplication, which we shall call a \textit{symmetry group}.

Note the setting in which we present symmetries here is stricter than the conventional definition, which considers any operator $S$ that commutes with the Hamiltonian, i.e. $[S, H_\mathcal{T}] = 0$, to be a symmetry. Such an operator need not commute with the individual terms as we require here. For example, in the fermionic picture, the number operator $\sum_i a_i^\dag a_i$ (where $a$ is the fermionic annihilation operator and its Hermitian conjugate $a^\dag$ represents the creation operator) commutes with the full second-quantized molecular Hamiltonian, but not with an arbitrary excitation term.

The operators of $\mathcal{S}$ will in general be algebraically dependent, but the theory of stabilizers \cite{gottesman1997stabilizer} ensures the existence of a set of independent generators $\mathcal{G}$ such that $\overline{\mathcal{S}} = \braket{\mathcal{G}}$. Now, recall the Clifford group consists of unitary operators $U \in \mathscr{B}(\mathscr{H})$ (meaning $UU^\dag = U^\dag U = \mathbbm{1}$) with the property $U \bm{\sigma} U^\dag \in \mathcal{P}_N \fa \bm{\sigma}\in\mathcal{P}_N$, i.e., $U$ \textit{normalizes} the Pauli group. We may construct a Clifford operation $U$ mapping each symmetry generator to distinct single-qubit Pauli operators $\sigma_p$, where we are free to choose $p \in \{1,2,3\}$. More precisely, there exists a subset of qubit positions $\I{stab} \subset \mathbb{Z}_N$ satisfying $|\I{stab}| = |\mathcal{G}|$ and a bijective map $f:\mathcal{G} \to \I{stab}$ such that 
\begin{equation}\label{unitary_rotations}
    U G U^\dag = \sigma^{(f(G))}_p \;\fa G \in \mathcal{G}.    
\end{equation}

This is a powerful concept that provides a mechanism for reducing the number of qubits in the Hamiltonian whilst preserving its energy spectrum. This is at the core of qubit tapering \cite{bravyi2017tapering, setia2020reducing}, in which it is observed that
\begin{equation}
    [G, H_\mathcal{T}] = 0 \implies [\sigma_p^{(f(G))}, H_\mathcal{T}^\prime] = 0 \;\fa G \in \mathcal{G},
\end{equation}
implying the rotated Hamiltonian $H_\mathcal{T}^\prime \coloneqq U H_\mathcal{T} U^\dag$ consists solely of identity or Pauli $\sigma_p$ operators in the qubit positions indexed by $\I{stab}$. Taking expectation values, one may replace the qubits $\I{stab}$ by their eigenvalues $ \nu_i = \pm1$; each assignment
\begin{equation}
    \bm{\nu} = (\nu_i)_{i \in \I{stab}} \in \{\pm1\}^{\times \I{stab}}
\end{equation}
defines a symmetry \textit{sector} and at least one such sector will contain the true solution to the eigenvalue problem. Note the other sectors still have physical significance and may for example relate to solutions with different particle numbers or to excited states. In the Supporting Information we report the symmetry generators and corresponding sector for the Hamiltonians representing the molecular systems listed in Table \protect\ref{mol_suite}.

A quantum state consistent with any such sector must be stabilized by the operators $\nu_i \sigma_p^{(i)}$ and we may define a projection onto the corresponding stabilizer subspace. In general, a projection is defined to be an \textit{idempotent} operator $P \in \mathscr{B}(\mathscr{H})$, i.e. $P^2 = P$; the projection onto the $\pm1$-eigenspace of a single-qubit Pauli operator $\sigma_p$ for $p \in \{1,2,3\}$ may be written
\begin{equation}\label{frob_cov}
    P_p^{\pm} \coloneqq \frac{1}{2}\big(I \pm \sigma_p\big).
\end{equation}
States with no component inside the chosen eigenspace are mapped to zero and observe that 
\begin{equation}\label{proj_action}
    P_p^{\pm} \sigma_q P_p^{\pm} = \pm\delta_{p,q} P_p^{\pm}
\end{equation}
for $q \in \{1,2,3\}$.

Let $\mathscr{H}_\mathrm{stab}$ be the reduced Hilbert space supported by the stabilized qubits $\I{stab}$ and $\mathscr{H}_\mathrm{red}$ its complement such that $\mathscr{H} = \mathscr{H}_\mathrm{stab} \otimes \mathscr{H}_\mathrm{red}$. Given an assignment of eigenvalues $\bm{\nu} \in \{\pm1\}^{\times \I{stab}}$, we may project onto the corresponding sector via 
\begin{equation}\label{nu_proj}
    P_{\bm{\nu}} \coloneqq \bigotimes_{i \in \I{stab}} P_p^{\nu_i}
\end{equation}
and subsequently perform a \textit{partial trace} over the stabilized qubits $\I{stab}$. This is effected by the unique linear map $\Tr_\mathrm{stab}: \mathscr{H} \to \mathscr{H}_\mathrm{red}$ satisfying the property $\Tr_\mathrm{stab}\big( A \otimes B \big) = \Tr\big(A\big)B$ for all $A \in \mathscr{B}(\mathscr{H}_\mathrm{stab})$ and $B \in \mathscr{B}(\mathscr{H}_\mathrm{red})$.

Finally, we may define the full stabilizer subspace projection map 
\begin{equation}\label{S3_proj}
\begin{aligned}
    \pi_{\bm{\nu}}^{U}:
    \mathscr{B}(\mathscr{H}) \to{} & \mathscr{B}(\mathscr{H}_\mathrm{red});\\ 
    A \mapsto{} & \Tr_\mathrm{stab} \big( P_{\bm{\nu}} U A U^\dag P_{\bm{\nu}} \big)
\end{aligned}
\end{equation}
which, using the linearity of $\Tr_\mathrm{stab}$, yields a reduced Hamiltonian
\begin{equation}\label{taper_ham}
\begin{aligned}
    H_\mathcal{T}^\mathrm{red} 
    \coloneqq{} & \pi_{\bm{\nu}}^{U} (H_\mathcal{T}) \\
    ={} & \sum_{\bm{\sigma} \in \mathcal{T}} h^\prime_{\bm{\sigma}} \bm{\sigma}^\prime_\mathrm{red}
\end{aligned}
\end{equation}
where $\bm{\sigma}^\prime = U \bm{\sigma} U^\dag = \bigotimes_{i=0}^{N-1} \sigma_{q_i}$ and we have written $\bm{\sigma}^\prime = \bm{\sigma}^\prime_\mathrm{stab} \otimes \bm{\sigma}^\prime_\mathrm{red}$. The new coefficients $h_{\bm{\sigma}}^\prime \coloneqq h_{\bm{\sigma}} \prod_{\substack{i \in \I{stab} \\ q_i \neq 0}} \nu_i$ differ from $h_{\bm{\sigma}}$ by a sign dependent on the chosen symmetry sector. 

In qubit tapering $U$ is taken as \eqref{unitary_rotations}, with the corresponding basis $\mathcal{G}$ a generating set for a full Hamiltonian symmetry \cite{bravyi2017tapering, setia2020reducing}. Assuming identification of the correct sector, the ground state energy of the $(N-|\mathcal{G}|)$-qubit reduced Hamiltonian $H_{\mathcal{T}}^\mathrm{red}$ will coincide with the true value of the full system $H_\mathcal{T}$.

%Note that, although the Clifford operations do not alter the number of terms of the Hamiltonian, its eigenstates will be transformed -- observing that $\braket{H_\mathcal{T}}_\psi = \braket{H_\mathcal{T}^\prime}_{U^\dag \psi}$, the quantum state $U^\dag \ket{\psi}$ will have $\mathcal{O}(2^{|\mathcal{G}|})$-times the number of basis states compared with $\ket{\psi}$ in its expansion. This has the effect of promoting otherwise negligible, low-lying excitations to prominence within the spectrum of $H^\prime_\mathcal{T}$. At first glance, one might expect that it is therefore not sufficient to consider only ans\"atze that express just the most significant excitations above some reference state, instead having to characterise those of low energy as well. However, this is not the case, since the subsequent projection will force certain eigenstates and thus discard others, resulting in no net change.

This stabilizer projection procedure is straightforward with respect to the Hamiltonian, since the stabilized qubits contain only operators with non-zero image under conjugation with $P_{\bm{\nu}}$. However, suppose we were to take another observable $A \in \mathscr{B}(\mathscr{H})$ and wish to determine a reduced form on $\mathscr{B}(\mathscr{H}_\mathrm{red})$ that is consistent with the reduced Hamiltonian $H_{\mathcal{T}}^\mathrm{red}$. This may be achieved by following precisely the same process that was applied to $H_\mathcal{T}$, but the symmetry $\mathcal{S}$ will not in general be a symmetry of $A$ and therefore the `symmetry-breaking' terms (those which anticommute with the generators $\mathcal{G}$) will vanish under projection onto the stabilizer subspace, as per \eqref{proj_action}. Letting $\mathcal{A} \subset \mathcal{P}_N$ be the set of terms in the Pauli-basis expansion of $A$, observe that
\begin{equation}\label{taper_obs}
\begin{aligned}
    A^\mathrm{red} 
    \coloneqq{} & \pi_{\bm{\nu}}^{U} (A) \\
    ={} & \sum_{\bm{\sigma} \in \mathcal{A}} h_{\bm{\sigma}} \Tr(P_{\bm{\nu}}\bm{\sigma}^\prime_\mathrm{stab} P_{\bm{\nu}}) \bm{\sigma}^\prime_\mathrm{red}\\
    ={} & \sum_{\bm{\sigma} \in \mathcal{A}} h_{\bm{\sigma}} \Tr \Big( \bigotimes_{\substack{i \in \I{stab} \\ q_i = 0}} P_p^{\nu_i} \; \otimes \\ & \hspace{1.96cm} \bigotimes_{\substack{i \in \I{stab} \\ q_i \neq 0}} \underbrace{P_p^{\nu_i} \sigma_{q_i} P_p^{\nu_i}}_{\;\;=\nu_i\delta_{p,q_i} P_p^{\nu_i}} \Big) \bm{\sigma}^\prime_\mathrm{red} \hspace{2cm} \\
    ={} & \sum_{\bm{\sigma} \in \mathcal{A}} h_{\bm{\sigma}} \bm{\sigma}_\mathrm{red}^\prime \underbrace{\Tr ( P_{\bm{\nu}})}_{=1} \prod_{\substack{i \in \I{stab} \\ q_i \neq 0}} \nu_i \delta_{p, q_i}\\
    ={} & \sum_{\substack{\bm{\sigma} \in \mathcal{A} \\ q_i \in \{0, p\} \\ \fa i \in \I{stab}}} h^\prime_{\bm{\sigma}} \bm{\sigma}_\mathrm{red}^\prime,
\end{aligned}
\end{equation}
recalling that $q_i$ indicates the type of single-qubit Pauli acting on qubit position $i \in \mathbb{Z}_N$ in some tensor product $\bm{\sigma}$, defined in Section \protect\ref{prelim}.

The resulting form is identical to \eqref{taper_ham}, except we are explicit that the terms surviving projection are only those whose qubit positions indexed by $\I{stab}$ consist exclusively of identity and Pauli $\sigma_p$ operators; this is trivially true for the Hamiltonian by construction. Most importantly, this extends the stabilizer subspace projection to ans\"atze defined on the full system for use in variational algorithms. It should be noted that the above operations are classically tractable and can be implemented efficiently in the symplectic representation of Pauli operators.

We would be remiss not to draw attention to the likeness of \eqref{S3_proj} with Positive Operator-Valued Measures (POVM) \cite{nielsen2010quantum}; indeed, the projectors \eqref{nu_proj} define a complete set of \textit{Kraus} operators \cite{kraus1983states}. The stabilizer subspace projection procedure is reduced to a matter of enforcing a partial measurement over some subsystem of the full problem, for which the relevant outcomes have been determined via an auxiliary method. For example, this could involve identifying a quantum state with a known non-zero overlap with the true ground state; measuring the symmetry generators $\mathcal{G}$ in this state will yield the correct sector.

Hartree-Fock often provides such a state for electronic structure problems, although it is not immune to failure; this is particularly true in the strongly-correlated regime. In these cases, we should defer to more effective reference states such as those obtained from Møller–Plesset perturbation theory (MP2), coupled-cluster (CC) methods and so on. One can imagine a hierarchy of increasingly precise ground state approximations, for which we should hope to obtain at some point a non-zero overlap with the true ground state.

\section{CS-VQE in the stabilizer formalism}\label{CS-VQE_stab}

We now describe the \textit{Contextual Subspace VQE} (CS-VQE) method in the stabilizer setting introduced in Section \protect\ref{stabilizers}. CS-VQE partitions the Hamiltonian \eqref{hamiltonian} into two disjoint components -- one that is noncontextual %and appears to be heuristically solvable by classical means, despite arbitrary instances being NP-complete \cite{kirby2019contextuality, kirby2020classical}, 
and another that is contextual which provides quantum corrections to the former via VQE \cite{kirby2021contextual}. Explicitly, this allows us to write
\begin{equation}\label{ham_decomp}
    H_{\mathcal{T}} = H_{\I[T]{nc}} + H_{\I[T]{c}}
\end{equation}
where $\I[T]{nc}$ is a noncontextual set of Pauli operators and $\I[T]{c} \coloneqq \mathcal{T} \setminus \I[T]{nc}$ is what remains, which will in general be contextual.

CS-VQE differs from qubit tapering (described in section \protect\ref{stabilizers}) in the following way: the latter exploits existing (i.e. physical) symmetries of the Hamiltonian, whereas in CS-VQE we impose additional `pseudo-symmetries' derived from the noncontextual Hamiltonian. This results in a loss of information, since any terms of $\mathcal{T}$ not commuting with the symmetry generators will vanish under projection. %but choosing these symmetries carefully one can still retain sufficient precision for the purposes of the eigenvalue problem of interest.

\subsection{The noncontextual problem}\label{nc_problem}

The notion of contextuality goes back to the Bell-Kochen-Specker theorem \cite{bell1964einstein, bell1966problem, kochen1975problem}. Here we use an explicit condition for the noncontextuality of a set of Pauli operators, developed by Kirby \& Love \cite{kirby2019contextuality} and independently by Raussendorf et al. \cite{raussendorf2020phase}. Strictly speaking, this condition tests for strong measurement contextuality. In this setting, a set $\mathcal{T}_\mathrm{nc}$ is understood to be noncontextual if and only if commutation forms an equivalence relation on $\mathcal{T}_\mathrm{nc} \setminus \mathcal{S}$, where we have defined the sub-Hamiltonian symmetry $\mathcal{S} \coloneqq \{\bm{\sigma} \in \mathcal{T}_\mathrm{nc} |\, [\bm{\sigma}, \bm{\sigma}^\prime]=0 \fa \bm{\sigma}^\prime \in \mathcal{T}_\mathrm{nc} \}$. There is an implied structure
\begin{equation}\label{noncon_decomp}
    \mathcal{T}_\mathrm{nc} = \mathcal{S} \cup \mathcal{C}_1 \cup \dots \cup \mathcal{C}_M,  
\end{equation}
where the $\mathcal{C}_i$ are equivalence classes with respect to commutation -- in other words, elements of the same class commute and across classes they anticommute. Conversely, such a set of Pauli operators is contextual if and only if commutation fails to be transitive on $\mathcal{T}_\mathrm{nc} \setminus \mathcal{S}$. 

The symmetry $\mathcal{S}$ can be expanded by taking pairwise products within equivalence classes -- since $\{C_i, C_j\} = 0$ for $C_i \in \mathcal{C}_i, C_j \in \mathcal{C}_j$ with $i \neq j$, it is the case that $[C_i C_i^\prime, C_j C_j^\prime] = 0$ and we may define $\mathcal{S}^\prime = \mathcal{S} \cup \bigcup_{i=1}^M \{C_i C_i^\prime |\; C_i, C_i^\prime \in \mathcal{C}_i\}$. As before, in Section \protect\ref{stabilizers}, $\mathcal{S}^\prime$ induces a symmetry group for which one may define independent generators $\mathcal{G}$ and a Clifford operation $U_\mathcal{G}$ mapping the generators to single-qubit Pauli operators; the expectation value over these qubits will again be determined by an assignment $\bm{\nu} \in \{\pm1\}^{\times |\mathcal{G}|}$ of eigenvalues, analogous to the selection of a symmetry sector in qubit tapering. 

From each equivalence class $\mathcal{C}_i$ we select a representative $C_i$ and construct an observable $C(\bm{r}) \coloneqq \sum_{i=1}^M r_i C_i$ where $\bm{r} \in \mathbb{R}^{M}$ and $|\bm{r}| = 1$. Kirby \& Love \cite{kirby2020classical} found that quantum states $\ket{\psi_{(\bm{\nu}, \bm{r})}} \in \mathscr{H}$ stabilized by the operators $\{\nu_{f(G)} G \;|\; G \in \mathcal{G}\} \cup \{C(\bm{r})\}$ are consistent with a classical objective function $\eta(\bm{\nu}, \bm{r})$ (derived in the Supporting Information), in the sense that $\eta(\bm{\nu}, \bm{r})$ coincides with the noncontextual energy expectation value $\braket{H_{\mathcal{T}_\mathrm{nc}}}_{\psi_{(\bm{\nu}, \bm{r})}}$ for all parametrizations $(\bm{\nu}, \bm{r})$. This is a consequence of the joint probability distribution chosen over the phase-space points of their (epistricted) model \cite{kirby2020classical, Spekkens2016}.

The noncontextual energy spectrum is therefore parametrized by two vectors: the $\pm1$ eigenvalue assignments $\bm{\nu}$, determining the contribution of the universally commuting terms, and $\bm{r}$, encapsulating the remaining pairwise anticommuting classes. In this sense, we may refer to $(\bm{\nu}, \bm{r})$ as a state of the noncontextual Hamiltonian itself, abstracted from quantum states of the corresponding stabilizer subspace. Optimizing over these parameters, we obtain the noncontextual ground state energy
\begin{equation}
    \epsilon_0^\mathrm{nc} \coloneqq \min_{\substack{\bm{\nu} \in \{\pm1\}^{\times |\mathcal{G}|} \\ \bm{r} \in \mathbb{R}^M : |\bm{r}|=1}} \eta(\bm{\nu}, \bm{r})
\end{equation}
and call an element $(\bm{\nu}, \bm{r})$ of the preimage $\eta^{-1}(\epsilon_0^\mathrm{nc})$ a noncontextual ground state of $H_{\I[T]{nc}}$. Let us denote by $\Delta_\mathrm{nc} \coloneqq |\epsilon_0^\mathrm{nc} - \epsilon_0|$ the absolute error with respect to the true ground state energy.

As a classical estimate to the ground state energy of the full Hamiltonian $H_\mathcal{T}$, in Section \protect\ref{simres} we found the difference between the noncontextual ground state and Hartree-Fock energy to be negligible for each of the molecules simulated, since the heuristic used to choose $H_{\I[T]{nc}}$ prioritizes diagonal Hamiltonian terms. In principle, it may be an improvement upon Hartree-Fock as the noncontextual set can also take into account an off-diagonal contribution within the anticommuting classes. This is highly dependent on the chosen form of noncontextual set; a reformulation in terms of graphs -- e.g. representing Pauli operators as nodes with (non)adjacency indicating (anti)commutation -- will allow one to identify what the equivalent problem(s) are in computer science and therefore draw upon the vast body of existing research and select the best algorithms designed to solve such computational problems of graph theory. It should be noted the `optimal' noncontextual subset will not necessarily be that which minimizes the noncontextual ground state energy and some consideration of the resulting quantum corrections must inform this choice.

\subsection{Quantum corrections}\label{corrections}

Our simulation approach has thus far been strictly classical -- now we arrive at the quantum element of CS-VQE. We have derived a classical estimate of the ground state energy from the noncontextual part of the Hamiltonian $H_{\I[T]{nc}}$; however, the contextual component $H_{\I[T]{c}}$ has so far been neglected. 

While $C(\bm{r})$ is not a stabilizer in the strict sense (it is not an element of the Pauli group), it is unitarily equivalent to one as a linear combination of anticommuting Pauli elements. Similar to the symmetry generators $\mathcal{G}$, it is possible to define a unitary operation $U_C$ mapping $C(\bm{r})$ onto a single-qubit Pauli operator, following the approach of unitary partitioning \cite{izmaylov2019unitary, zhao2020measurement, bonet2020nearly, ralli2021implementation, ralli2022unitary}. However, unlike the $U_\mathcal{G}$ rotation, $U_\mathcal{C}$ is not Clifford as it collapses $M$ terms onto a single Pauli operator and can therefore introduce additional terms to the Hamiltonian. Kirby et al. \cite{kirby2021contextual} cautioned that, in principle, this increase in Hamiltonian complexity could be exponential in the number of equivalence classes $M$, namely a scaling of $\mathcal{O}(2^M)$. However, Ralli et al. \cite{ralli2022unitary} demonstrated that the general scaling is $\mathcal{O}(x^{M-1})$ where $x \in [1,2]$; that is, still exponential, yet the necessary conditions to obtain the worst-case $x=2$ are contrived and have not been observed for any molecular Hamiltonians investigated to date. Regardless, one may circumvent this potential adverse scaling by implementing the linear combination of unitaries approach at the expense of one ancillary qubit and its necessarily probabilistic nature \cite{zhao2020measurement, ralli2021implementation, ralli2022unitary}.

Appending $C(\bm{r})$ to our set of generators $\Tilde{\mathcal{G}} \coloneqq \mathcal{G} \cup \{C(\bm{r})\}$ and defining $U \coloneqq U_C U_\mathcal{G}$, there exists a subset of qubit indices $\I{stab}$ satisfying $|\I{stab}|=|\Tilde{\mathcal{G}}|$ and a bijective map $f:\Tilde{\mathcal{G}} \to \I{stab}$ such that $U G U^\dag = \sigma_p^{(f(G))}$ for each $G \in \mathcal{\Tilde{G}}$. We reiterate that $p \in \{1, 2, 3\}$ may be chosen at will; the approach taken by Kirby et al. \cite{kirby2021contextual} is to select $p=3$ to enforce diagonal generators.

Suppose we have a quantum state $\ket{\psi_{(\bm{\nu}, \bm{r})}}$ that is consistent with $\epsilon_0^\mathrm{nc}$; since the rotated state $\ket{\psi^\prime_{(\bm{\nu}, \bm{r})}} = U \ket{\psi_{(\bm{\nu}, \bm{r})}}$ must be stabilized by $\sigma_p^{(i)} \fa i \in \I{stab}$, the qubit positions $\I{stab}$ must be fixed. This implies a decomposition 
\begin{equation}
    \ket{\psi^\prime_{(\bm{\nu}, \bm{r})}} = \ket{b_{(\bm{\nu}, \bm{r})}}_\mathrm{stab} \otimes \ket{\varphi}_\mathrm{red}    
\end{equation}
where $\ket{b_{(\bm{\nu}, \bm{r})}}$ represents a single basis state of $\mathscr{H}_\mathrm{stab}$ and $\ket{\varphi} \in \mathscr{H}_\mathrm{red}$ is independent of the parameters $(\bm{\nu}, \bm{r})$. Therefore, the expectation value of the full Hamiltonian may be expressed as
\begin{equation}\label{expect_proj}
    \braket{H_\mathcal{T}}_{\psi_{(\bm{\nu}, \bm{r})}} = \epsilon_0^\mathrm{nc} + \braket{\pi_{\bm{\nu}}^{U} (H_{\mathcal{T}_\mathrm{c}}) }_{\varphi},
\end{equation}
where $\pi^{U}_{\bm{\nu}} (H_{\mathcal{T}_\mathrm{c}})$ contains only the terms of the contextual Hamiltonian that commute with all the noncontextual generators, just as in \eqref{taper_obs}. It was observed by Kirby et al. \cite{kirby2021contextual} that any term which anticommutes with at least one noncontextual generator must have zero expectation value and our stabilizer subspace projection captures this fact.

Inspecting \eqref{expect_proj}, we may optimize freely over quantum states $\varphi$, i.e., we are not constrained by the noncontextual ground state within $\mathscr{H}_\mathrm{red}$. In fact, we may absorb the noncontextual ground state energy into the reduced contextual Hamiltonian, defining the \textit{contextual subspace Hamiltonian} 
\begin{equation}\label{first_contextual}
    \Tilde{H}_{\mathcal{T}_c} \coloneqq \epsilon_0^\mathrm{nc} \cdot \mathbbm{1} + \pi_{\bm{\nu}}^{U} (H_{\mathcal{T}_\mathrm{c}});
\end{equation}
this form is obtained naturally when applying the stabilizer subspace projection to the full Hamiltonian, which automatically includes the noncontextual energy by fixing the corresponding eigenvalue assignments.

Now, we may perform unconstrained VQE to obtain a quantum-corrected estimate 
\begin{equation}
    \epsilon_0^\mathrm{c} \coloneqq \min_{\ket{\varphi} \in \mathscr{H}_\mathrm{red}} \braket{\Tilde{H}_{\mathcal{T}_\mathrm{c}}}_\varphi
\end{equation}
of the true ground state energy with absolute error $\Delta_\mathrm{c} \coloneqq |\epsilon_0^\mathrm{c} - \epsilon_0| \leq \Delta_\mathrm{nc}$. We have equality when the stabilizers span every qubit position, which is the case when $|\Tilde{\mathcal{G}}| = N$ since the generators must be algebraically independent: this means the initial quantum correction is trivial as the noncontextual part determines the entire system.

For instances of the electronic structure problem there is no guarantee that $\epsilon_0^\mathrm{c}$ will achieve chemical accuracy ($\Delta_\mathrm{c} < 1.6\mathrm{mHa} \approx 4\mathrm{kJ/mol}$) and, indeed, it might not improve upon the noncontextual estimate (although it will never be worse, due to the variational principle applying in this case). However, one can easily define a subset of $\I[T]{nc}$ that is again noncontextual -- this is achieved by discarding one of the noncontextual generators $G \in \Tilde{\mathcal{G}}$, along with the operators that it generates. We now append the discarded operators to the contextual Hamiltonian, relaxing the stabilizer constraint on the qubit position $f(G)$ and permitting a search over its Hilbert space. This process may be iterated until the noncontextual set is exhausted and we recover full VQE. This means that, unless the ground state energy of $H_{\I[T]{nc}}$ and $H$ coincides, CS-VQE will improve upon the noncontextual energy using less quantum resource than full VQE -- this is more rigorously defined in the next section.

In summary, what we have described here is a technique of scaling the relative sizes of the noncontextual (read classical) and contextual (read quantum) simulations in a reciprocal manner. We can therefore trade-off quantum and classical workloads in CS-VQE.

\subsection{Expanding the contextual subspace}\label{expanding}

Now we describe the process of growing the contextual subspace more rigorously. We select a subset of noncontextual generators $\mathcal{F} \subset \Tilde{\mathcal{G}}$ whose stabilizer constraints we mean to enforce and construct a new noncontextual set $\I[T]{nc}^\prime \coloneqq \I[T]{nc} \cap \overline{\mathcal{F}}$; the contextual set is expanded accordingly by appending the terms not generated by $\mathcal{F}$, i.e., $\I[T]{c}^\prime \coloneqq \I[T]{c} \cup (\I[T]{nc} \setminus \overline{\mathcal{F}})$. As before, there exists a unitary operation $U_{\mathcal{F}}$, a subset of qubit indices $\I{fix} \subset \I{stab}$ and a bijective map $f:\mathcal{F} \to \I{fix}$ satisfying $U_\mathcal{F} G U_\mathcal{F}^\dag = \sigma_p^{(f(G))} \;\fa G \in \mathcal{F}$ (the rotation $U_\mathcal{F}$ may or may not be Clifford depending on whether $C(\bm{r})$ is among the stabilizers we wish to fix). 

Denote by $\epsilon_0^\mathrm{nc}(\mathcal{F})$ the ground state energy of the new noncontextual Hamiltonian $\I[T]{nc}^\prime$ with absolute error $\Delta_\mathrm{nc}(\mathcal{F}) \geq \Delta_\mathrm{nc}$. While this is weaker as an estimate of the true ground state energy of the full system, at the very least we are guaranteed to recover the initial noncontextual ground state energy from performing a simulation of the expanded contextual subspace \cite{kirby2021contextual}, which we describe below. 

The stabilizer constraints of $\mathcal{F}$ are enforced over the Hilbert space $\mathscr{H}_\mathrm{fix} = (\mathbb{C}^2)^{\otimes \I{fix}}$ of qubits indexed by $\I{fix}$, whereas we may perform a VQE simulation over $\mathscr{H}_\mathrm{sim} = (\mathbb{C}^2)^{\otimes \I{sim}}$, the Hilbert space of the remaining $N-|\mathcal{F}|$ qubits indexed by $\I{sim} = \mathbb{Z}_N \setminus \I{fix}$. Invoking the stabilizer subspace projection map $\pi_{\bm{\nu}^\prime}^{U_\mathcal{F}}$%: \mathscr{B}(\mathscr{H}) \to \mathscr{B}(\mathscr{H}_\mathrm{sim})$ 
with the eigenvalue assignments $\bm{\nu}^\prime = (\nu_i)_{i \in \I{fix}}$ yields an expanded contextual subspace Hamiltonian
\begin{equation}\label{gen_contextual_ham}
    \Tilde{H}_{\mathcal{T}^\prime_c} \coloneqq \epsilon_0^\mathrm{nc}(\mathcal{F}) \cdot \mathbbm{1} + \pi_{\bm{\nu}^\prime}^{U_\mathcal{F}} (H_{\mathcal{T}^\prime_\mathrm{c}}).
\end{equation}
Performing an $|\I{sim}|$-qubit VQE simulation over the contextual subspace we obtain a new quantum-corrected estimate
\begin{equation}
    \epsilon_0^\mathrm{c}(\mathcal{F}) \coloneqq \min_{\ket{\varphi} \in \mathscr{H}_\mathrm{sim}} \braket{ \Tilde{H}_{\mathcal{T}^\prime_c}}_\varphi
\end{equation}
with an error satisfying $\Delta_\mathrm{c}(\mathcal{F}) \leq \Delta_\mathrm{c}$. Recall that $\Delta_\mathrm{c} = \Delta_\mathrm{c}(\Tilde{\mathcal{G}})$ corresponds with the contextual error when we enforce the full set of noncontextual stabilizers.

Observe that, when $|\I{sim}| = N$, we are simply performing full VQE over the entire system -- this occurs when we do not enforce the stabilizer constraint for any of the noncontextual generators, i.e. $\mathcal{F} = \emptyset$. Therefore, it must be the case that
\begin{equation}\label{target_error}
    \Delta_\mathrm{c}(\mathcal{F}) \to 0 \;\; \text{as} \;\; |\mathcal{F}| \to 0.
\end{equation}
Furthermore, given a nested sequence of generator subsets $(\mathcal{F}_i)_i$ with $\mathcal{F}_{i+1} \subset \mathcal{F}_{i}$, then $\Delta_\mathrm{c}(\mathcal{F}_{i+1}) \leq \Delta_\mathrm{c}(\mathcal{F}_{i})$ and the convergence is monotonic. In this way, CS-VQE describes an interpolation between a purely classical estimate of the ground state energy and a full VQE simulation of the Hamiltonian performed over some ansatz space. In the context of electronic structure calculations, this often permits one to achieve chemical accuracy at a saving of qubit resource, as indicated by Kirby et al. \cite{kirby2021contextual} for a suite of tapered test molecules of up to 18 qubits.

Suppose we wish to find the optimal contextual subspace Hamiltonian of size $N^\prime < N$. The problem reduces to minimizing the error $\Delta_\mathrm{c}(\mathcal{F})$ over the $|\Tilde{\mathcal{G}}| \choose N-N^\prime$ generator subsets $\mathcal{F} \subset \Tilde{\mathcal{G}}$ satisfying $|\mathcal{F}| = N - N^\prime$. CS-VQE is highly sensitive to this choice and remains a vital open question for the continued success of the technique. For chemistry applications, we grow the contextual subspace until the CS-VQE error attains chemical accuracy, which means finding the minimal $\mathcal{F}$ such that $\Delta_\mathrm{c}(\mathcal{F}) < 1.6\mathrm{mHa}$. In general, we will not have access to a target energy and so will not necessarily know when the desired precision is achieved; instead, we might iterate until the VQE convergence is within some fixed bound.

Greedily selecting combinations of $d \leq N$ generators that yield the greatest reduction in error, necessitating $\sum_{k=0}^{\lfloor N/d \rfloor} {N-dk \choose d} = \mathcal{O}(N^{d+1})$ CS-VQE simulations, is an effective stabilizer relaxation ordering heuristic. Taking $d=2$ produces a good balance between efficiency and efficacy \cite{kirby2021contextual}, but there is room for more targeted approaches that exploit some structure of the underlying problem. For example, in quantum chemistry problems it could be that one should relax the stabilizers that have non-trivial action near the Fermi level -- between the highest occupied molecular orbital (HOMO) and lowest unoccupied molecular orbital (LUMO). Excitations clustered around this gap are more likely to appear in the true ground state and should therefore not be assigned definite values under the noncontextual projection. This idea comes from the theory of pseudopotential approximations \cite{schwerdtfeger2011pseudopotential}, in which it is observed that chemically relevant electrons are predominantly those of the valence space, whereas the core may be `frozen', thus reducing the electronic complexity.

Alternatively, one might define a Hamiltonian term-importance metric that considers coefficient magnitudes \cite{wecker2014gate} or second-order response with respect to a perturbation of the Hartree-Fock state \cite{poulin2014trotter}. In relation to this, it is also not clear which features of a molecular system mean that it might be more or less amenable to CS-VQE; additional insight here would allow one to predict how many qubits will be required to simulate a given problem to chemical accuracy.

\subsection{The noncontextual projection ansatz}

CS-VQE has thus far not been applied to systems exceeding 18 qubits and the resulting reduced Hamiltonians \eqref{gen_contextual_ham} have been solved by direct diagonalization \cite{kirby2021contextual} -- clearly, this will not scale to larger systems, with the required classical memory increasing exponentially. Instead, they must be simulated by performing VQE routines, but defining an ansatz for the contextual subspace provided an obstacle to achieving this in practice.

However, having now placed the problem within the stabilizer formalism described in Section \protect\ref{stabilizers}, we have already introduced (in Sections \protect\ref{nc_problem} - \protect\ref{expanding}) the tools necessary to restrict an ansatz of the form \eqref{vanilla_anz} -- defined over the full system -- to the contextual subspace \eqref{gen_contextual_ham}. The approach adopted here is equivalent to that which we defined for qubit tapering in \eqref{taper_obs}. To restrict a parametrized ansatz operator 
\begin{equation}
    A(\bm{\theta}) = \sum_{\bm{\sigma} \in \mathcal{A}} \theta_{\bm{\sigma}} \bm{\sigma} \mapsto \Tilde{A}(\bm{\theta}) \in \mathscr{B}(\mathscr{H}_\mathrm{sim})    
\end{equation}
in line with the stabilizer constraints $\mathcal{F} \subset \Tilde{\mathcal{G}}$ we may simply call upon the stabilizer subspace projection map $\pi_{\bm{\nu}^\prime}^{U_\mathcal{F}}$ once more, which yields a restricted ansatz state
\begin{equation}\label{proj_anz}
    \ket{\Tilde{\psi}_\mathrm{anz}(\bm{\theta})} \coloneqq e^{i \tilde{A}(\bm{\theta})} \ket{\Tilde{\psi}_\mathrm{ref}} \in \mathscr{H}_\mathrm{sim}
\end{equation}
where 
\begin{equation}
    \Tilde{A}(\bm{\theta}) \coloneqq \pi_{\bm{\nu}^\prime}^{U_\mathcal{F}}\big(A(\bm{\theta})\big).
\end{equation}
Any rotated ansatz term $U_\mathcal{F} \bm{\sigma} U_\mathcal{F}^\dag$ that is not identity or a Pauli $\sigma_p$ on some subset of the qubit positions indexed by $\I{fix}$ will vanish.

The restricted reference state $\ket{\Tilde{\psi}_\mathrm{ref}}$ is obtained from a partial projective measurement of $U \ket{\psi_\mathrm{ref}}$ (see the discussion on POVMs in Section \protect\ref{stabilizers}) with outcomes defined by $\bm{\nu}^\prime$, which yields a product state
\begin{equation}\label{proj_ref}
    \frac{P_{\bm{\nu}^\prime} U \ket{\psi_\mathrm{ref}}}{\sqrt{\bra{\psi_\mathrm{ref}} U^\dag P_{\bm{\nu}^\prime} U \ket{\psi_\mathrm{ref}}}} = \ket{b_{(\bm{\nu}, \bm{r})}}_{\mathrm{fix}} \otimes \ket{\Tilde{\psi}_\mathrm{ref}}_{\mathrm{sim}}.
\end{equation}
The post-measurement state $\ket{b_{(\bm{\nu}, \bm{r})}} \in \mathscr{H}_\mathrm{fix}$ on the noncontextual subspace represents a single basis vector and can therefore be disregarded, leaving just the state of the contextual subspace -- this we take as reference for our restricted ansatz. We stress this `measurement' is not performed in circuit but is instead to be evaluated classically when constructing the restricted ansatz circuit.

We may now define the contextual subspace energy expectation function
\begin{equation}\label{E_cs}
    \Tilde{E}(\bm{\theta}) \coloneqq \braket{\Tilde{H}_{\I[T]{c}^\prime}}_{\Tilde{\psi}_\mathrm{anz}(\bm{\theta})}
\end{equation}
with $\Tilde{H}_{\I[T]{c}^\prime}$ as in \eqref{gen_contextual_ham}, at which point we have reduced the problem to standard VQE, performed over a subspace of the full problem. 

%Chemically, one might interpret this in the following way: the stabilizer constraint over $\I{fix}$ enforces the occupancy/vacancy of these qubits in accordance with the noncontextual ground state and any operator exciting from/to the corresponding spin-orbitals should therefore be disallowed. This is somewhat conceptually reminiscent of frozen-core approximations, although the stabilizer rotations obscure this slightly.

In order to prepare the projected ansatz state \eqref{proj_anz}, we first initialize the $|\I{sim}|$-qubit quantum circuit in the noncontextual ground state, achieved by applying a Pauli $\sigma_1$ operator in each of the qubit positions $i \in \I{sim}$ such that $\nu_i = -1$. This is visible in Figure \protect\ref{fig:QLM_conv_results}, in which the VQE routine is initiated with the optimization parameters zeroed, i.e. $\bm{\theta} = \bm{0}$, and since $e^{i \Tilde{A}(\bm{0})} = \mathbbm{1}$ optimization begins at the noncontextual ground state energy.

It is not in general possible to implement the unitary operation $e^{i \Tilde{A}(\bm{\theta})}$ exactly as a quantum circuit (except for in the case of completely commuting terms $\mathcal{A}$ of $A(\bm{r})$), however one may do so approximately via the commonly used technique of Trotterization (see the Supporting Information for further details).

%Now that we have obtained an ansatz for the contextual subspace, the final obstacle to CS-VQE being deployed on NISQ hardware has been overcome. In the next section we...

\section{Simulation results}\label{simres}

The molecular systems that were simulated to benchmark the noncontextual projection ansatz for CS-VQE are given in Table \protect\ref{mol_suite}. The molecule geometries were obtained from the Computational Chemistry Comparison and Benchmark Database (CCCBDB) \cite{CCCBDB} and their Hamiltonians constructed using IBM's Qiskit Nature \cite{Qiskit} with PySCF the underlying quantum chemistry package \cite{sun2018pyscf}. %Our CS-VQE ansatz circuits were constructed using the functionality of Qiskit, whereas an Atos Quantum Learning Machine (QLM) provided by the Leibniz Supercomputing Centre (LRZ) facilitated compilation and execution of the quantum circuits.

\begin{table}[ht]
    \centering
    \begin{tabularx}{\linewidth}{@{}lYYYYr@{}} \toprule
    \multicolumn{3}{c}{Molecular systems} & \multicolumn{3}{c}{\hspace{5mm}Number of qubits} \\
    \cmidrule(r{-.5pt}){1-3} \cmidrule(l{10pt}){4-6}
    Name       & Charge& Mult.        & Full      & Taper      & {CS-VQE}  \\ \midrule
     \ce{Be}   & 0     &  1           & 10        & 5          & 3       \\
     \ce{B}    & 0     &  2           & 10        & 5          & 3       \\
     \ce{LiH}  & 0     &  1           & 12        & 8          & 4       \\
     \ce{BeH}  & +1    &  1           & 12        & 8          & 6       \\
     \ce{HF}   & 0     &  1           & 12        & 8          & 4       \\
     \ce{BeH2} & 0     &  1           & 14        & 9          & 7       \\
     \ce{H2O}  & 0     &  1           & 14        & 10         & 7       \\
     \ce{F2}   & 0     &  1           & 20        & 16         & 10      \\
     \ce{HCl}  & 0     &  1           & 20        & 17         & 4       \\
    \bottomrule
    \end{tabularx}
    \caption{The systems investigated to benchmark the noncontextual projection ansatz (all in the STO-3G basis). The CS-VQE column indicates the fewest number of qubits required to achieve chemical accuracy.}
    \label{mol_suite}
\end{table}

Before we evaluate the efficacy of our noncontextual projection ansatz, there are a few features of \eqref{proj_anz} that should be highlighted. First of all, in \eqref{proj_ref} we are applying the operation $U$ in-circuit, introducing further gates that will contribute additional noise. However, when the reference state is taken to be that of Hartree-Fock, we observed $U\psi_\mathrm{ref}$ to coincide with the noncontextual ground state. This is an artifact of the noncontextual set construction heuristic -- used within both this work and \cite{kirby2021contextual} -- prioritizing diagonal entries. This need not always be the case, but for the molecular systems investigated this allows us to avoid performing $U$ in-circuit and instead take the noncontextual ground state as our reference.

Secondly, application of the unitary partitioning rotations $U_C$ to the ansatz operator $A(\bm{\theta})$ may introduce additional terms by a scaling factor of $\mathcal{O}(x^{M-1})$ where $M$ is the number of equivalence classes in \eqref{noncon_decomp} and $x \in [1,2]$ a parameter depending on the given Hamiltonian, as discussed in section \ref{corrections}. We obtained $M=2$ for all of the molecules tested, although for a general Hamiltonian this need not be the case and is also dependent on the form of the noncontextual set $\I[T]{nc}$. Here we prioritize universally commuting terms, but it is equally valid to maximize the anticommuting contribution.

\begin{figure}[b!]
\centering

    \begin{subfigure}[t]{.32\linewidth}
    \includegraphics[width=\linewidth]{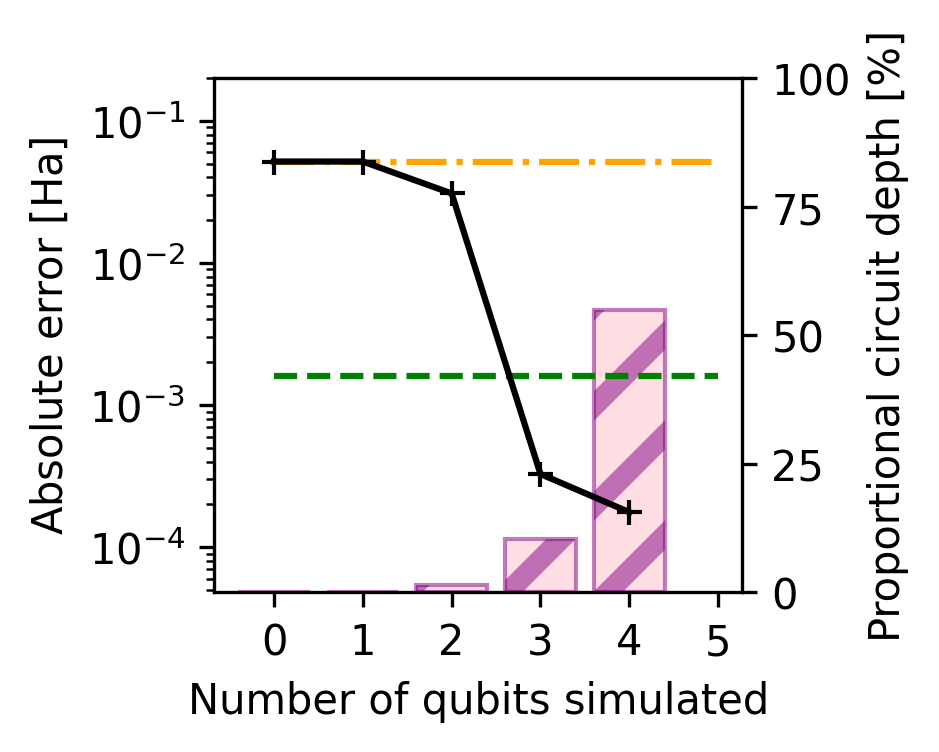}
    \caption{\ce{Be}}\label{fig:Be_3q_circ}
    \end{subfigure}
    \begin{subfigure}[t]{.32\linewidth}
    \includegraphics[width=\linewidth]{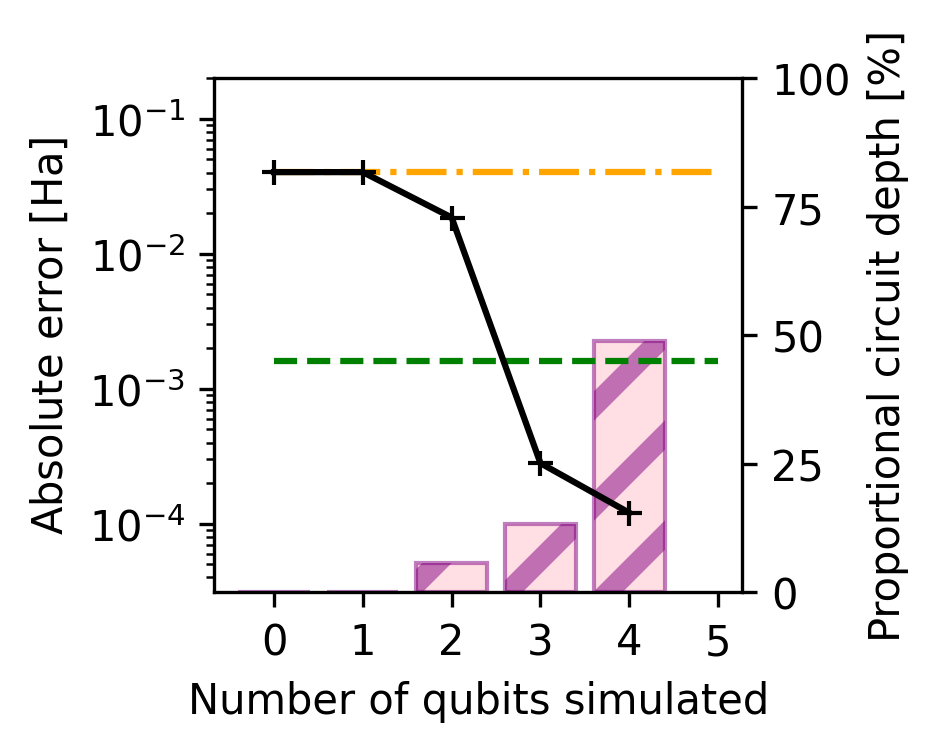}
    \caption{\ce{B}}\label{fig:B_3q_circ}
    \end{subfigure}
    \begin{subfigure}[t]{.32\linewidth}
    \includegraphics[width=\linewidth]{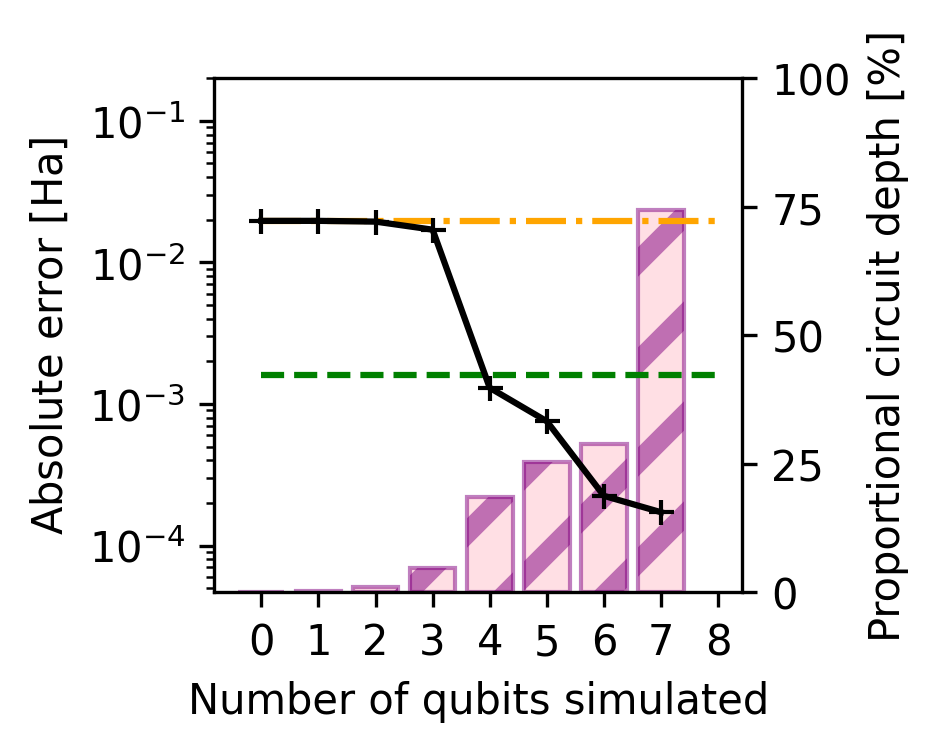}
    \caption{\ce{LiH}}\label{fig:LiH_4q_circ}
    \end{subfigure}
    \begin{subfigure}[t]{.32\linewidth}
    \includegraphics[width=\linewidth]{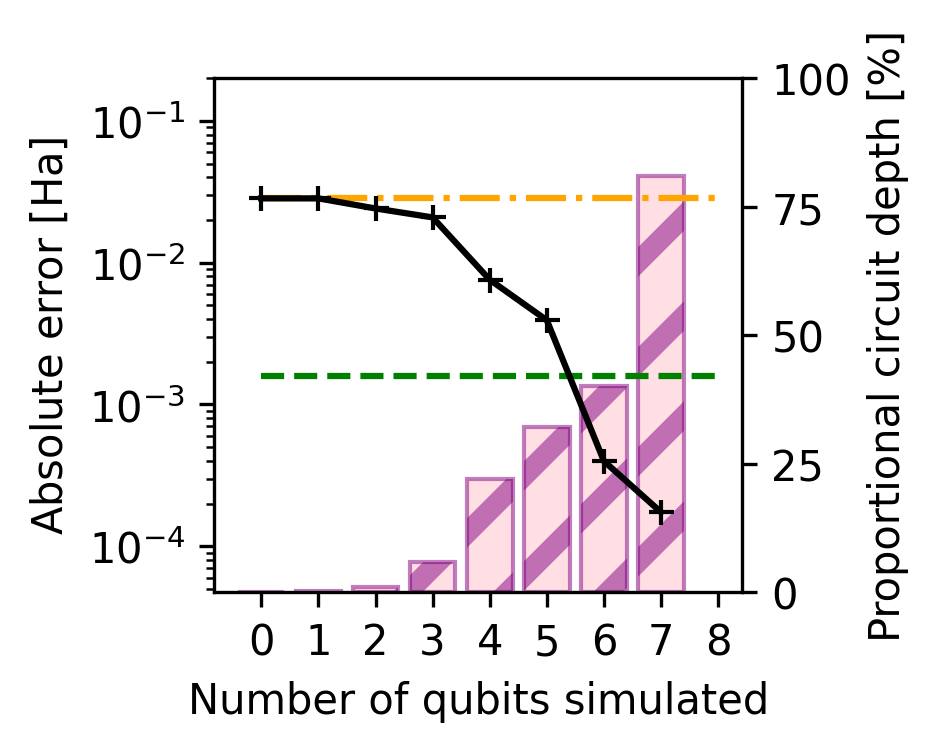}
    \caption{\ce{BeH+}}\label{fig:BeH+_6q_circ}
    \end{subfigure}
    \begin{subfigure}[t]{.32\linewidth}
    \includegraphics[width=\linewidth]{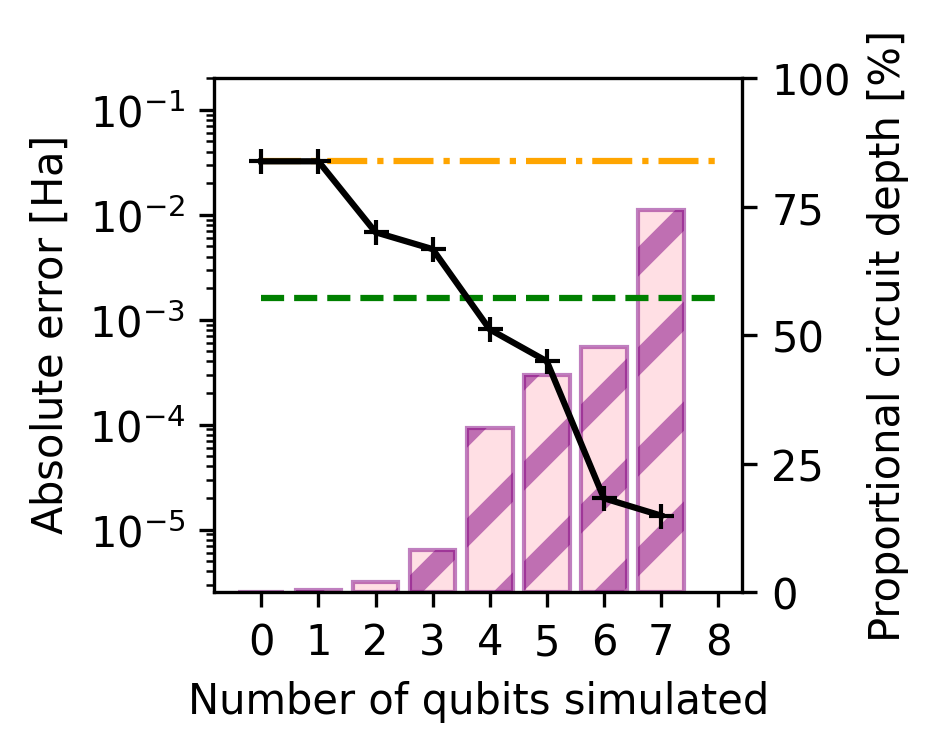}
    \caption{\ce{HF}}\label{fig:HF_4q_circ}
    \end{subfigure}
    \begin{subfigure}[t]{.32\linewidth}
    \includegraphics[width=\linewidth]{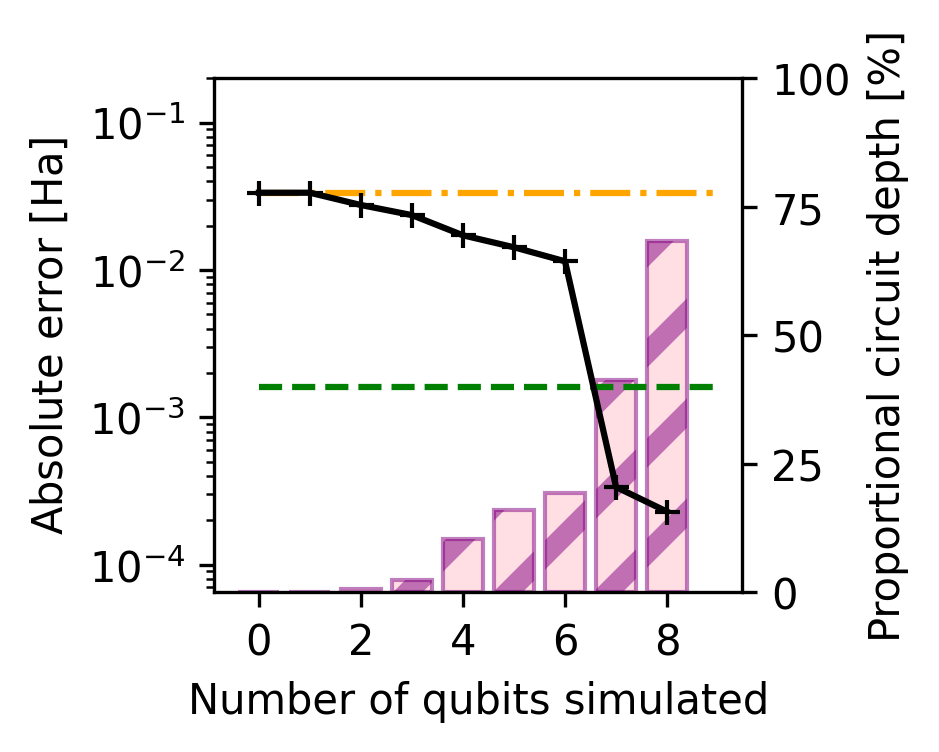}
    \caption{\ce{BeH2}}\label{fig:BeH2_7q_circ}
    \end{subfigure}
    \begin{subfigure}[t]{.32\linewidth}
    \includegraphics[width=\linewidth]{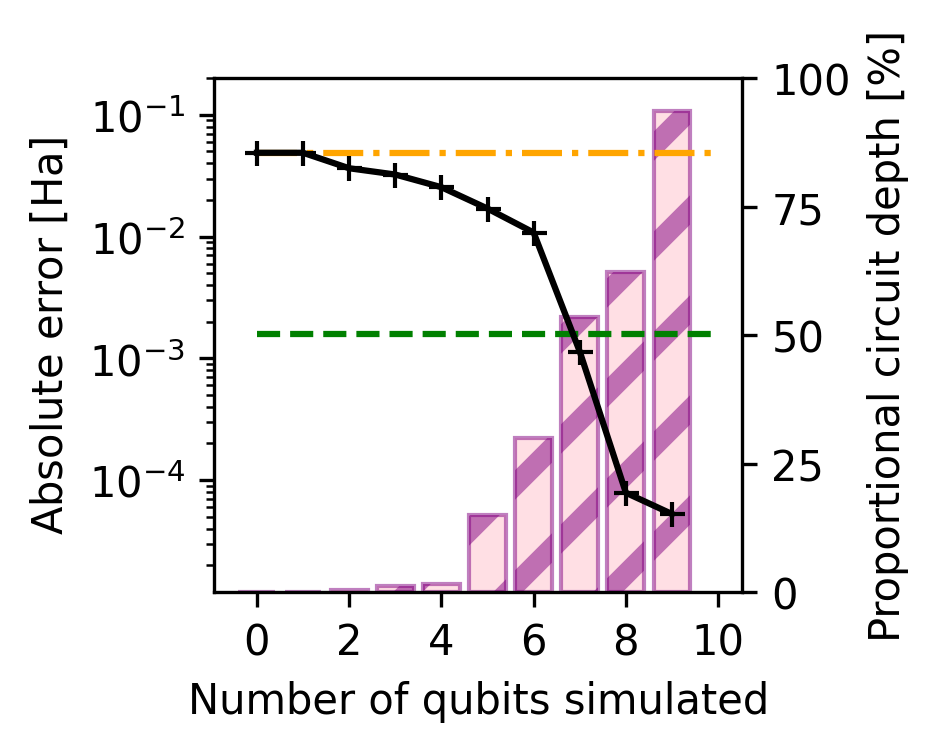}
    \caption{\ce{H2O}}\label{fig:H2O_7q_circ}
    \end{subfigure}
    \begin{subfigure}[t]{.32\linewidth}
    \includegraphics[width=\linewidth]{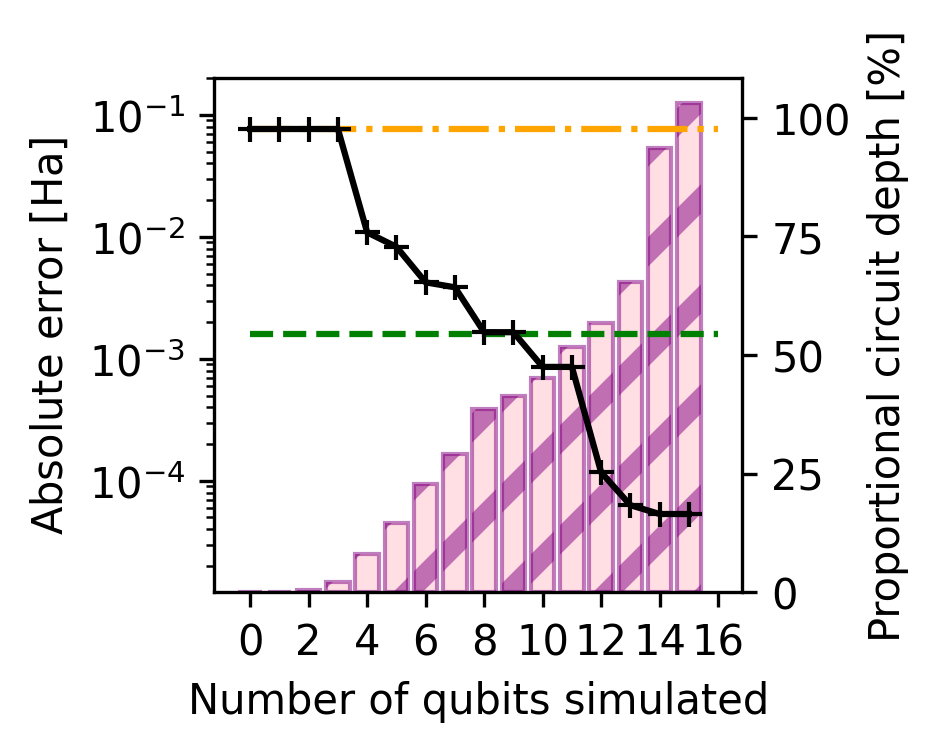}
    \caption{\ce{F2}}\label{fig:F2_10q_circ}
    \end{subfigure}
    \begin{subfigure}[t]{.32\linewidth}
    \includegraphics[width=\linewidth]{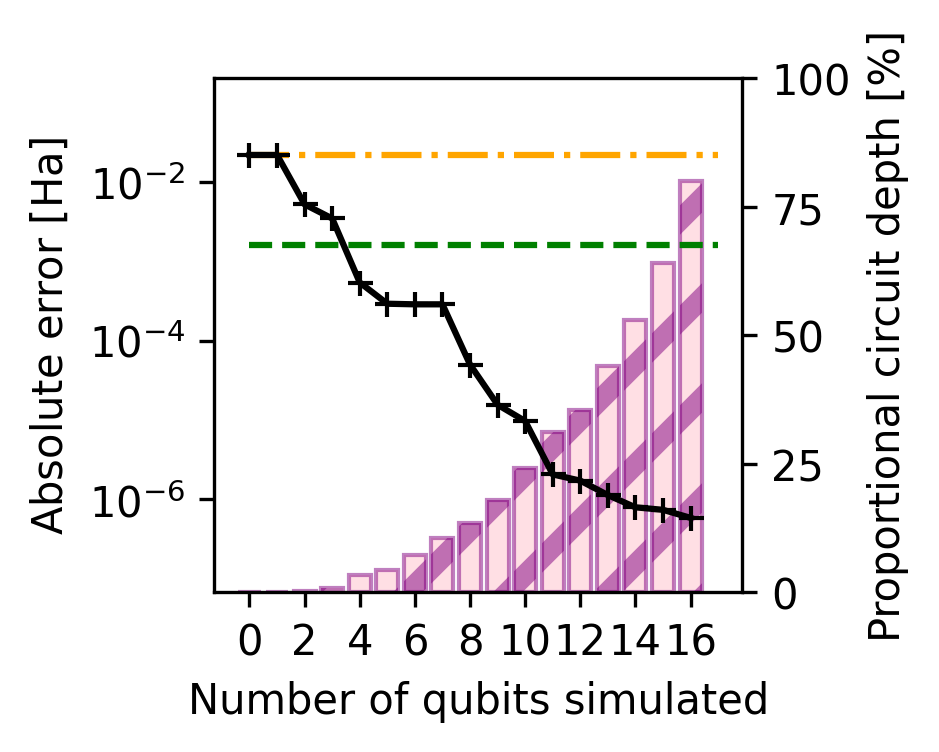}
    \caption{\ce{HCl}}\label{fig:HCl_4q_circ}
    \end{subfigure}
    \begin{subfigure}[b]{\linewidth}
    \includegraphics[width=\linewidth]{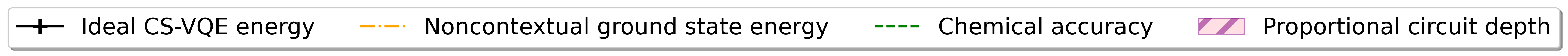}
    \end{subfigure}

\caption{Ideal CS-VQE errors (left-hand axis) and corresponding noncontextual projection ansatz circuit depths as a proportion of the full UCCSD operator from which it is derived (right-hand axis) against the number of qubits simulated.}
\label{fig:anz_depth}

\end{figure}

Despite this, upon the subsequent projection of $A(\bm{\theta})$, it is possible that a significant number of terms will vanish. This is highly dependent on the quality of the initial ansatz and how heavily it is supported on the stabilized qubit positions $\I{fix}$. Figure \protect\ref{fig:anz_depth} presents circuit depths of the noncontextual projection ansatz as a proportion of the base ansatz from which it is derived, in this case the unitary coupled-cluster singles and doubles (UCCSD) operator. A net reduction in circuit depth is observed, which is quite dramatic up to the point of reaching chemical accuracy in the CS-VQE routine; in Table \ref{mol_ansatze} we give the specific number of ansatz terms before and after application of the noncontextual projection to UCCSD and UCCSDT for the fewest number of qubits permitting chemical accuracy.

\begin{table}[h]
    %\centering
    \begin{tabularx}{\linewidth}{@{}llYYY@{}} \toprule
    & & \multicolumn{3}{c}{Number of terms in ansatz operator} \\
    \cmidrule(l){3-5}
    Molecule   & $|\I{sim}|$ &UCCSDT (full/proj) & UCCSD (full/proj) & ADAPT-VQE \\ \midrule
     \ce{Be}   & 3           & (48/6)            & (48/6)            & 5     \\
     \ce{B}    & 3           & (48/12)           & (32/4)            & 3     \\
     \ce{LiH}  & 4           & (704/53)          & (192/53)          & 5     \\
     \ce{BeH+} & 6           & (646/191)         & (166/79)          & 11    \\
     \ce{HF}   & 4           & (92/57)           & (92/57)           & 4     \\
     \ce{BeH2} & 7           & (1312/352)        & (224/96)          & 10    \\
     \ce{H2O}  & 7           & (1892/942)        & (324/238)         & 21    \\
     \ce{F2}   & 10          & (176/114)         & (176/114)         & 12    \\
     \ce{HCl}  & 4           & (348/40)          & (348/40)          & 4     \\
    \bottomrule
    \end{tabularx}
    \caption{The number of Pauli terms $|\mathcal{A}|$ for a selection of (tapered) ans\"atze. The $|\I{sim}|$ column indicates the number of qubits in the contextual subspace over which the ansatz is projected and each tuple (full/proj) gives the number of terms pre and post projection. The final column gives the number of ADAPT-VQE cycles required to achieve chemical accuracy, with the operator pool consisting of the projected UCCSD terms; each simulation is plotted in Figure \ref{fig:QLM_conv_results}.}
    \label{mol_ansatze}
\end{table}

\newpage
In order to identify a compact ansatz that closely captures the underlying chemistry with minimal redundancy, we employ the ADAPT-VQE methodology \cite{grimsley2019adaptive, tang2021qubit, shkolnikov2021avoiding, fedorov2021unitary}. The algorithm centres around an operator pool from which terms are selected in line with a gradient-based argument and appended to a dynamically expanding ansatz whose parameters are optimized at each cycle via VQE. The particular approach we implement here is that of qubit-ADAPT-VQE \cite{tang2021qubit}, which searches at the level of Jordan-Wigner encoded Pauli operators; the seminal ADAPT-VQE paper \cite{grimsley2019adaptive} instead defines its operator pool over fermionic excitations.

The Jordan-Wigner transformation \cite{jordan1993paulische} maps a single fermionic annihilation operator onto two Pauli operators
\begin{equation}
    a_i \mapsto \frac{1}{2}(\sigma_1^{(i)} + i \sigma_2^{(i)}) \otimes \bigotimes_{j<i} \sigma_3^{(j)},
\end{equation}
with the creation operator given by its Hermitian conjugate $a_i^\dag$. Therefore, an excitation on $s \in \mathbb{N}$ spin orbitals of the form
\begin{equation}
    \bm{a} = a_{i_1}^\dag \dots a_{i_s}^\dag a_{j_1} \dots a_{j_s}
\end{equation}
is represented by $2^{2s}$ Pauli operators under this encoding. In the unitary coupled cluster theory, we are interested rather in the operator $\bm{a} - \bm{a}^\dag$ to ensure unitarity upon exponentiation -- this may be expressed by $2^{2s-1}$ Pauli terms.

As such, after a mapping onto qubits via the Jordan-Wigner transformation, single, double and triple excitations account for 2, 8 and 32 Pauli operator terms respectively; while these are required to enforce various electronic symmetries in the ansatz state, not all are necessary to reach chemical accuracy. This idea lies behind qubit-ADAPT-VQE, which will select only the necessary Pauli terms and therefore yields considerably reduced circuit depths \cite{tang2021qubit}. 

To leverage ADAPT-VQE in the context of CS-VQE, we define an operator pool $\mathcal{O} \subset \mathcal{P}_N$ and apply to it the stabilizer subspace projection \eqref{S3_proj} to define a reduced pool $\pi_{\bm{\nu}^\prime}^{U_\mathcal{F}}(\mathcal{O})$ for the corresponding contextual subspace. The algorithm is then executed as normal, only terminating once the ADAPT-VQE energy is chemically accurate with respect to the FCI energy; for scalability, one should terminate computation when the largest gradient in magnitude falls below some predefined threshold, since the true ground state energy will not in general be known. In the Supporting Information, we provide a detailed description of the specific ADAPT-VQE implementation used within this work.

\begin{figure}[p]
\centering

    \begin{subfigure}[t]{.32\linewidth}
    \includegraphics[width=\linewidth]{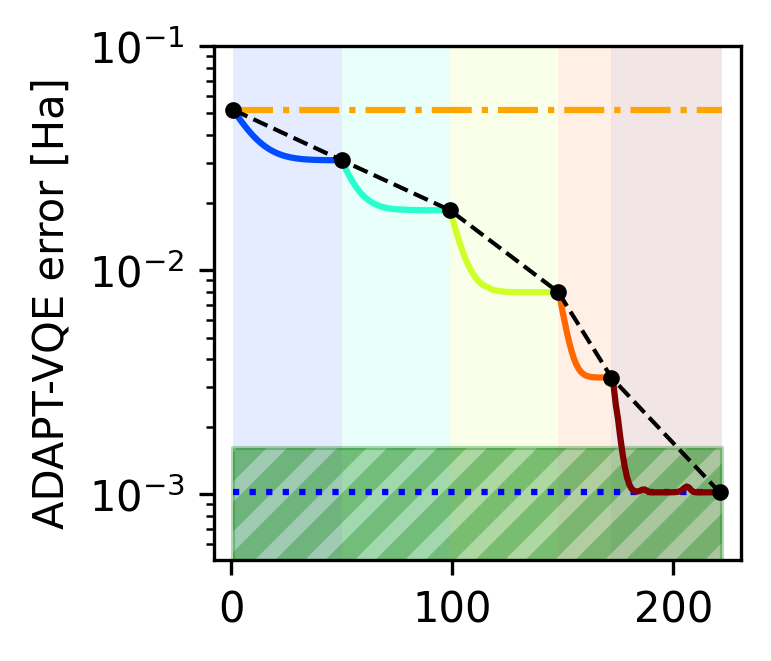}
    \caption{\ce{Be} 3-qubit CS-VQE}\label{fig:Be_3q_conv}
    \end{subfigure}
    \begin{subfigure}[t]{.32\linewidth}
    \includegraphics[width=\linewidth]{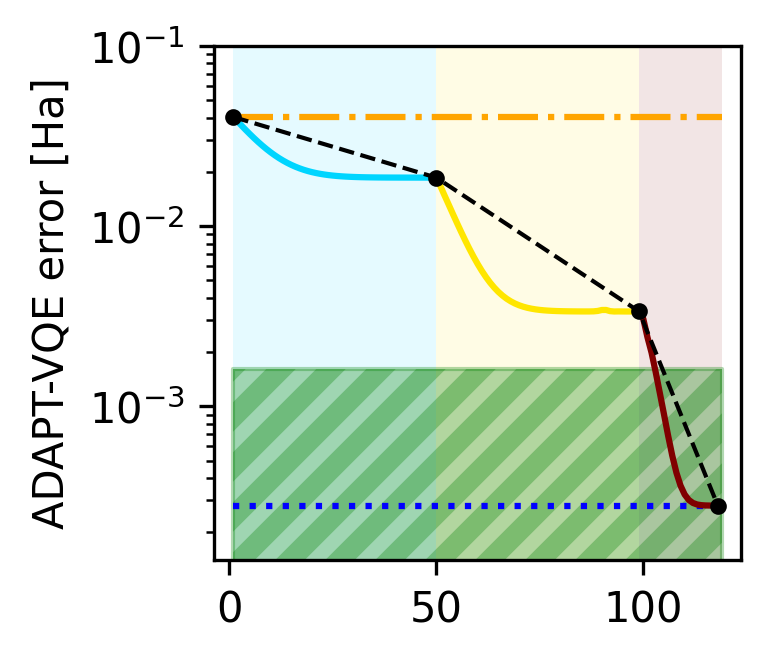}
    \caption{\ce{B} 3-qubit CS-VQE}\label{fig:B_3q_conv}
    \end{subfigure}
    \begin{subfigure}[t]{.32\linewidth}
    \includegraphics[width=\linewidth]{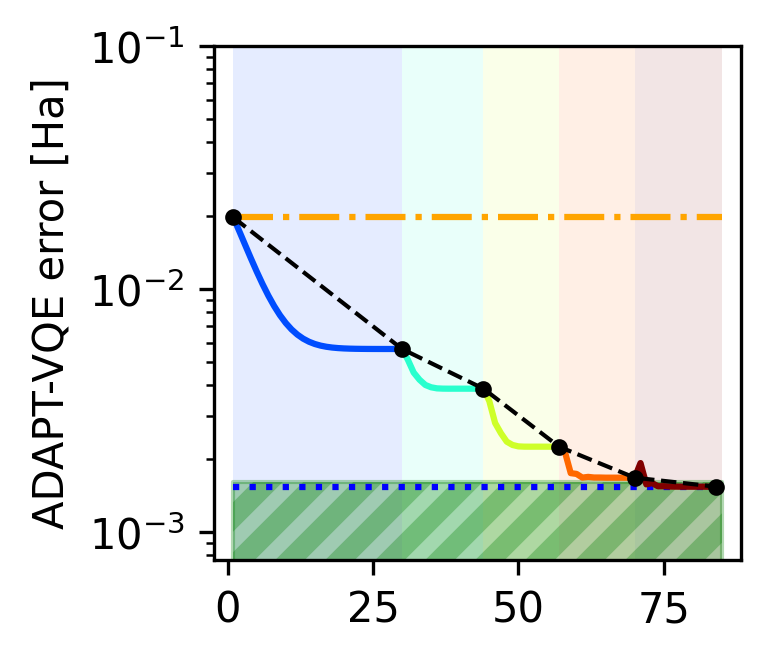}
    \caption{\ce{LiH} 4-qubit CS-VQE}\label{fig:LiH_4q_conv}
    \end{subfigure}
    \begin{subfigure}[t]{.32\linewidth}
    \includegraphics[width=\linewidth]{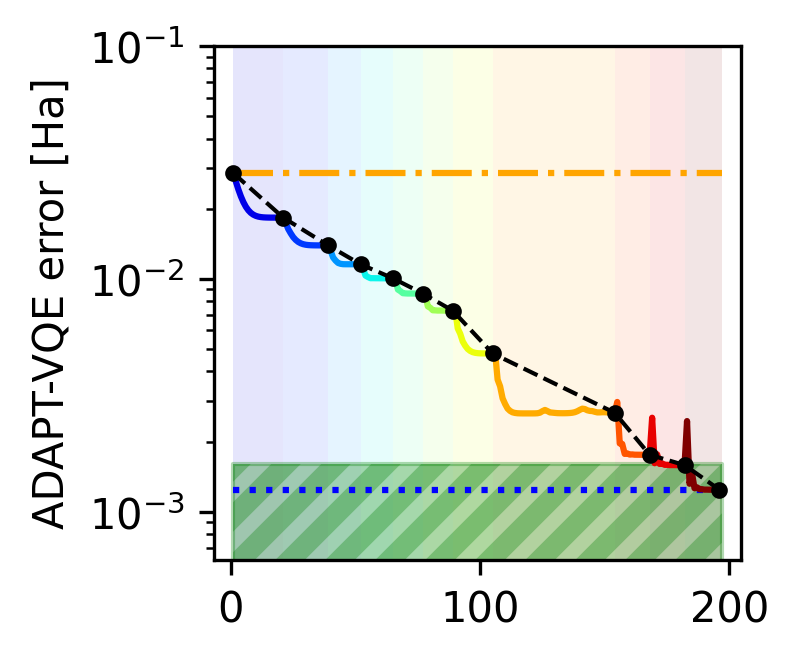}
    \caption{\ce{BeH+} 6-qubit CS-VQE}\label{fig:BeH+_6q_conv}
    \end{subfigure}
    \begin{subfigure}[t]{.32\linewidth}
    \includegraphics[width=\linewidth]{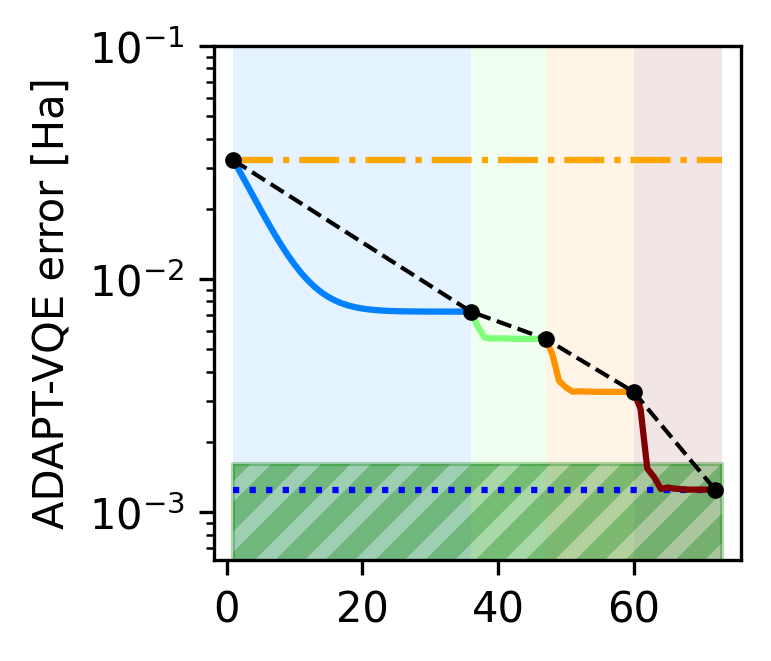}
    \caption{\ce{HF} 4-qubit CS-VQE}\label{fig:HF_4q_conv}
    \end{subfigure}
    \begin{subfigure}[t]{.32\linewidth}
    \includegraphics[width=\linewidth]{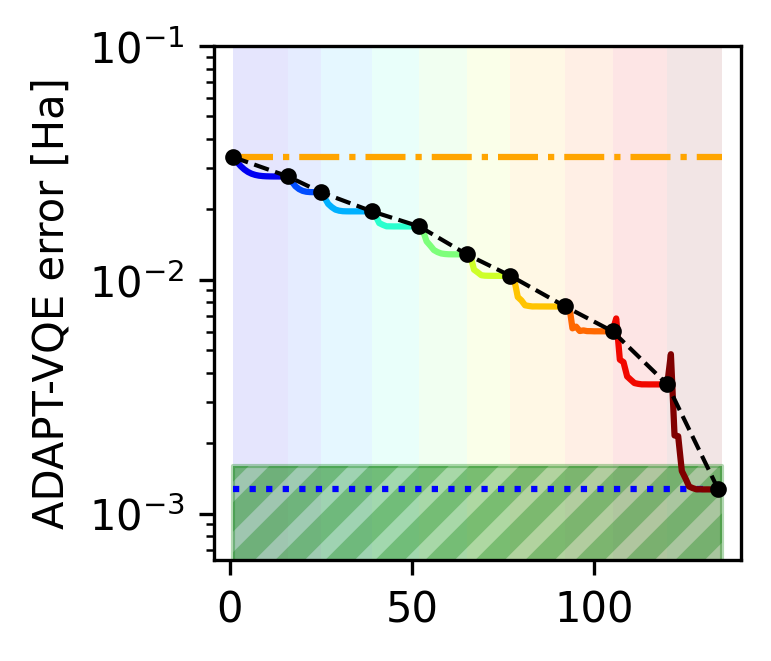}
    \caption{\ce{BeH2} 7-qubit CS-VQE}\label{fig:BeH2_7q_conv}
    \end{subfigure}
    \begin{subfigure}[t]{.32\linewidth}
    \includegraphics[width=\linewidth]{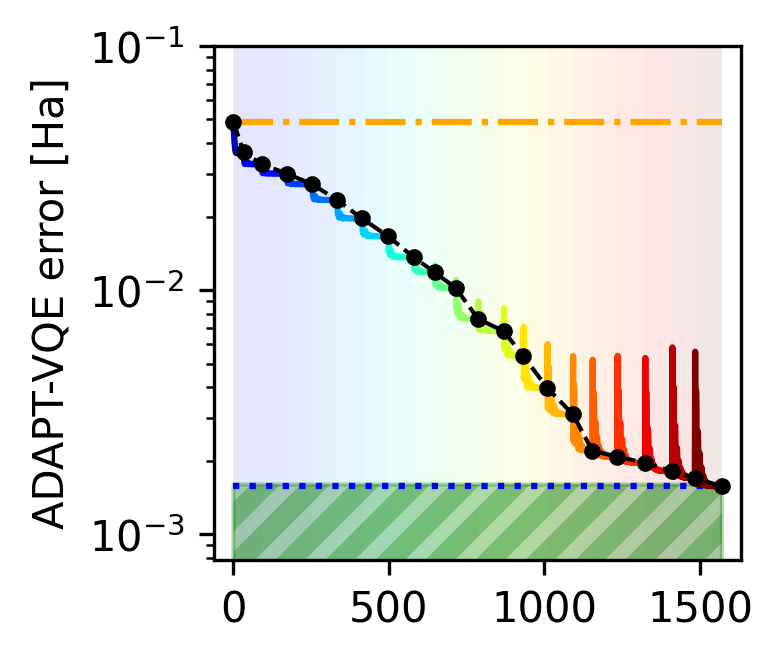}
    \caption{\ce{H2O} 7-qubit CS-VQE}\label{fig:H2O_7q_conv}
    \end{subfigure}
    \begin{subfigure}[t]{.32\linewidth}
    \includegraphics[width=\linewidth]{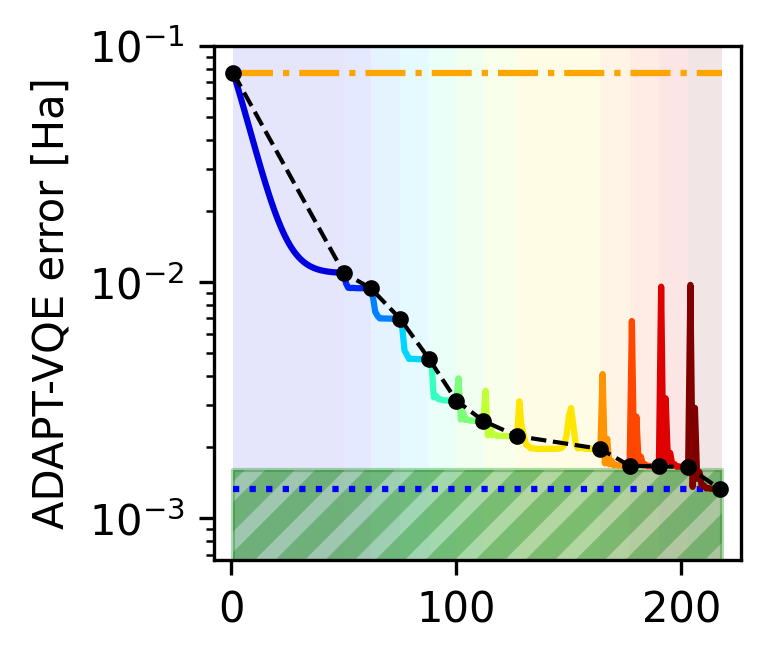}
    \caption{\ce{F2} 10-qubit CS-VQE}\label{fig:F2_10q_conv}
    \end{subfigure}
    \begin{subfigure}[t]{.32\linewidth}
    \includegraphics[width=\linewidth]{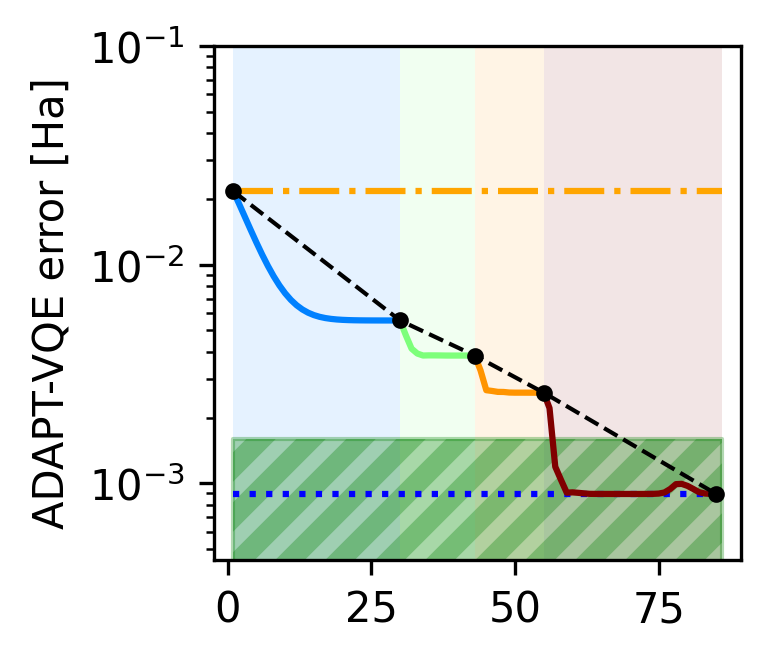}
    \caption{\ce{HCl} 4-qubit CS-VQE}\label{fig:HCl_4q_conv}
    \end{subfigure}
    \par\medskip
    \begin{subfigure}[b]{\linewidth}
    \includegraphics[width=\linewidth]{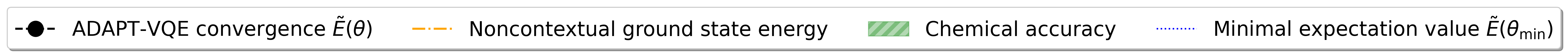}
    \end{subfigure}

\caption{Validation of the noncontextual projection approach to ansatz construction for CS-VQE \eqref{proj_anz}, used here in conjunction with ADAPT-VQE \cite{grimsley2019adaptive, tang2021qubit, shkolnikov2021avoiding, fedorov2021unitary}. We plot (on a $\log_{10}$ scale) the absolute error of wavefunction simulations conducted for the suite of trial molecules outlined in Table \protect\ref{mol_suite}, each shown to achieve chemical accuracy; the horizontal axis indicates the number of function evaluations (nfev). Adaptive moment estimation (Adam) \cite{kingma2014adam} is the classical optimizer taken in the VQE routine performed over the contextual subspace at each ADAPT-VQE cycle. The parameter gradients $\partial \Tilde{E}(\bm{\theta}) / \partial \theta_i$, required for both operator pool term selection and VQE, were computed using the parameter shift rule \protect{\cite{parrish2019hybrid}}.}
\label{fig:QLM_conv_results}
%\vspace{0.28cm}

\end{figure}

For the following, we take our pool $\mathcal{O}$ to be the terms of the UCCSD operator for each of the molecules in Table \ref{mol_suite} before tapering and projecting into the relevant contextual subspace. In Figure \protect\ref{fig:QLM_conv_results}, we present the ADAPT-VQE convergence data with expectation values obtained via exact wavefunction (statevector) calculations (i.e. no statistical/hardware noise); chemical accuracy is achieved in each instance. We used the adaptive moment estimation (Adam) \cite{kingma2014adam} classical optimizer and computed parameter gradients as per the parameter shift rule \cite{parrish2019hybrid}.

The number of ADAPT-VQE cycles (and therefore the number of terms in the resulting ansatz operator) are presented In Table \ref{mol_ansatze}, alongside the size of the projected UCCSD operator pool used; one observes a significant reduction in the number of terms. The optimized ADAPT-VQE ansatz operators are reported in the Supporting Information, along with a description of the smallest CS-VQE problem permitting chemical accuracy. This includes the optimal noncontextual generator subset $\mathcal{F}$, the resulting noncontextual projection ansatz \eqref{proj_anz}, restricted reference state $\ket{\Tilde{\psi}_\mathrm{ref}}$ \eqref{proj_ref}, the target error $\Delta_\mathrm{c}(\mathcal{F})$ \eqref{target_error} and that which was actually achieved in our VQE simulations (Figure \protect\ref{fig:QLM_conv_results}). We also include the corresponding contextual subspace Hamiltonians for reproducibility.

Extracting the optimal parameter configuration $\bm{\theta}_\mathrm{min}$ -- i.e. that which minimizes \eqref{E_cs} -- from the wavefunction simulations in Figure \protect\ref{fig:QLM_conv_results}, we subsequently assess the effect of sampling noise on the simulation error with our ansatz circuit preparing the optimal quantum state $\ket{\Tilde{\psi}_\mathrm{anz}(\bm{\theta}_\mathrm{min})}$. Note that, for each of the molecular systems in \ref{mol_suite}, $\bm{\theta}_\mathrm{min}$ is given explicitly in the Supporting Information.

%This is the more realistic situation for the NISQ era, whereby we perform many prepare-and-measure cycles, or shots, each time obtaining a binary output which corresponds with an expectation value for a single term of the Hamiltonian. Repeating this process term-wise builds a distribution of basis states and averaging over expectation values yields an estimate for the energy expectation of the Hamiltonian that is \textit{unbiased}, meaning we should expect errors to cancel in the asymptotic limit and produce good approximations of the energy.

\begin{figure}[p]
\centering

    \begin{subfigure}[t]{.31\linewidth}
    \includegraphics[width=\linewidth]{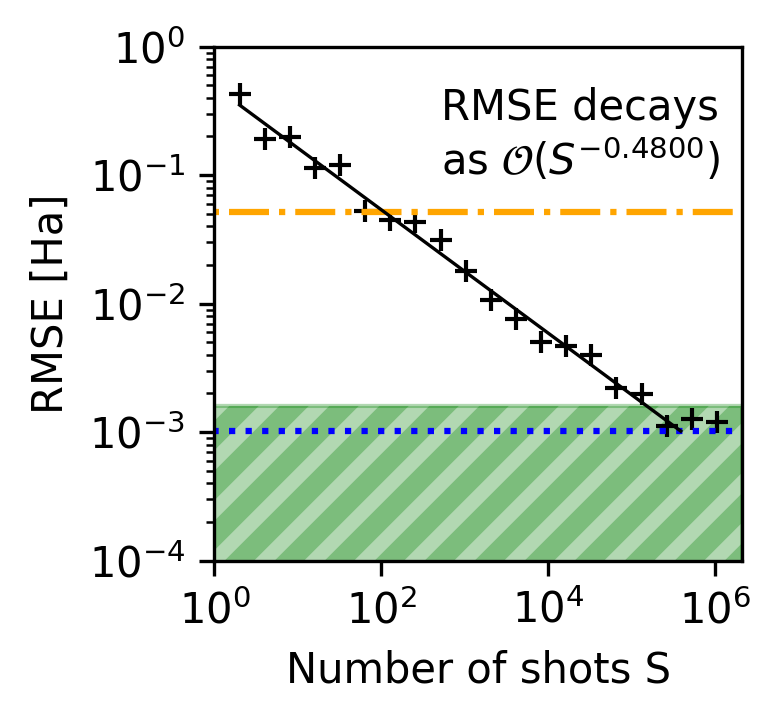}
    \caption{\ce{Be} 3-qubit CS-VQE}\label{fig:Be_3q_shot}
    \end{subfigure}
    \begin{subfigure}[t]{.31\linewidth}
    \includegraphics[width=\linewidth]{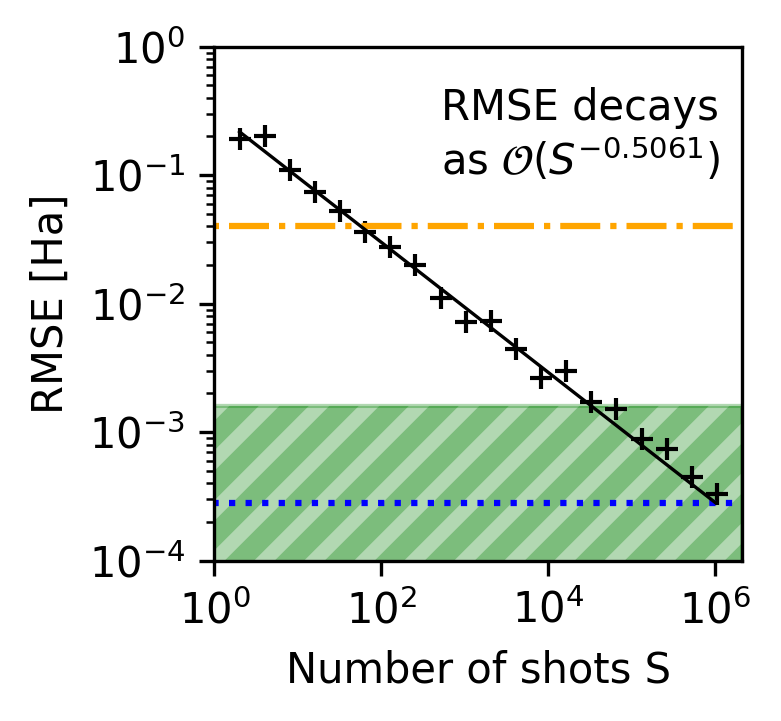}
    \caption{\ce{B} 3-qubit CS-VQE}\label{fig:B_3q_shot}
    \end{subfigure}
    \begin{subfigure}[t]{.31\linewidth}
    \includegraphics[width=\linewidth]{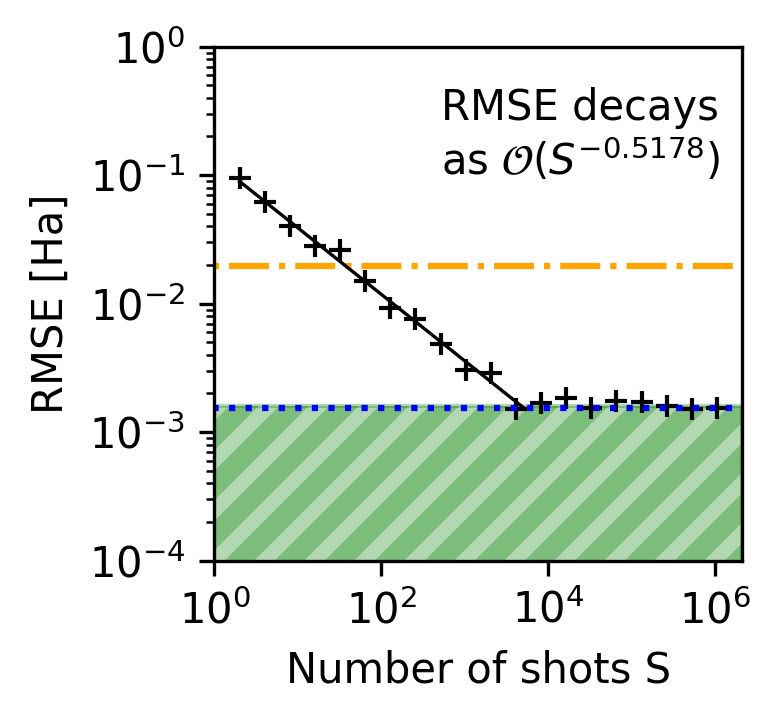}
    \caption{\ce{LiH} 4-qubit CS-VQE}\label{fig:LiH_4q_shot}
    \end{subfigure}
    \begin{subfigure}[t]{.31\linewidth}
    \includegraphics[width=\linewidth]{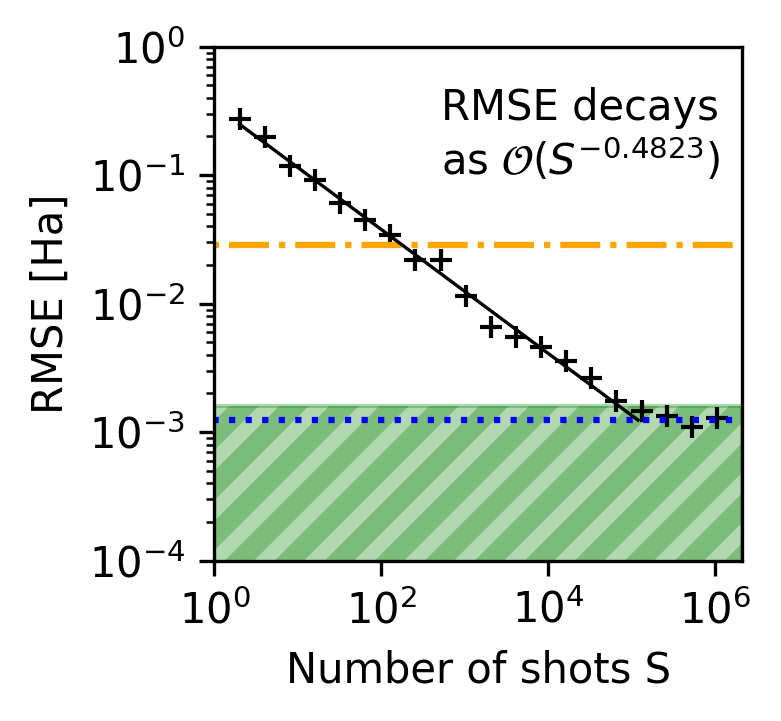}
    \caption{\ce{BeH+} 6-qubit CS-VQE}\label{fig:BeH+_6q_shot}
    \end{subfigure}
    \begin{subfigure}[t]{.31\linewidth}
    \includegraphics[width=\linewidth]{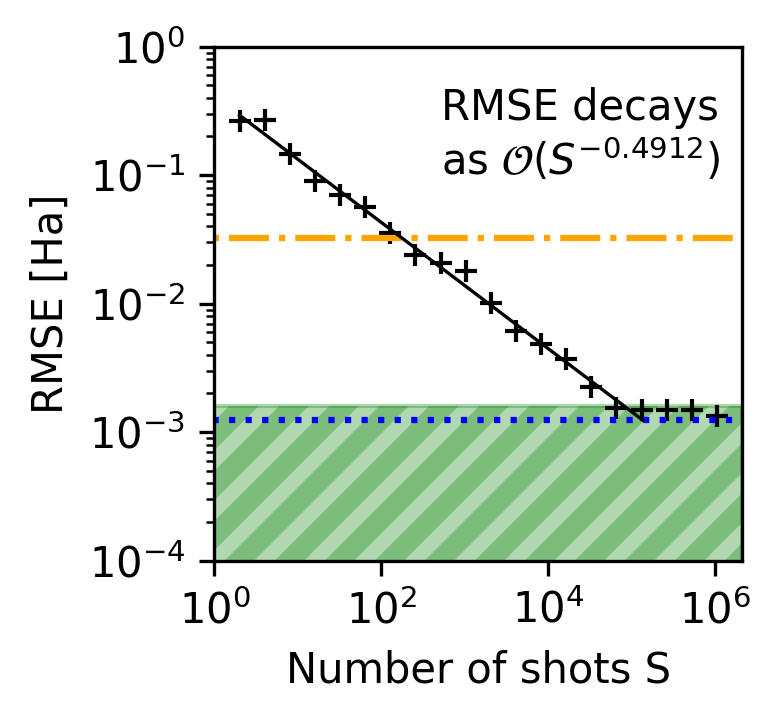}
    \caption{\ce{HF} 4-qubit CS-VQE}\label{fig:HF_4q_shot}
    \end{subfigure}
    \begin{subfigure}[t]{.31\linewidth}
    \includegraphics[width=\linewidth]{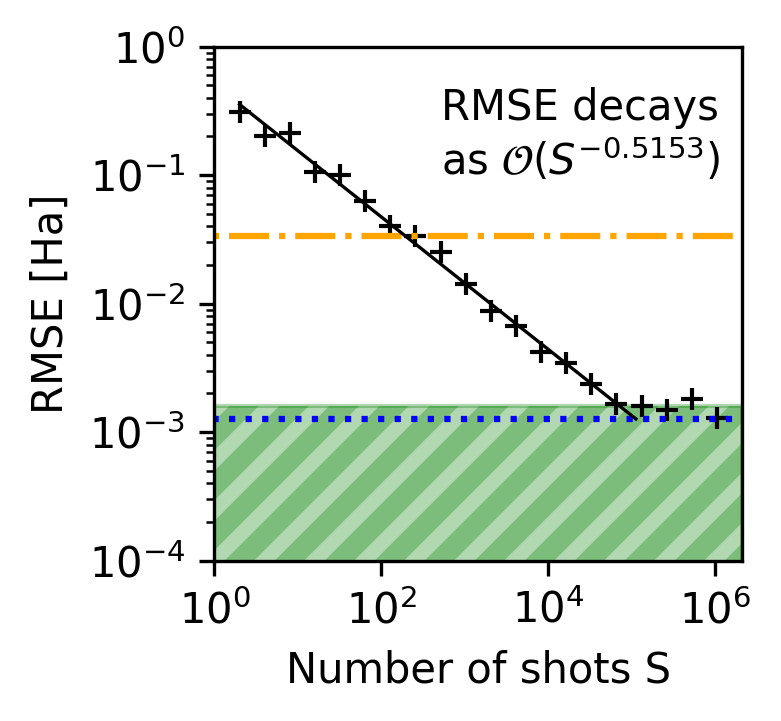}
    \caption{\ce{BeH2} 7-qubit CS-VQE}\label{fig:BeH2_7q_shot}
    \end{subfigure}
    \begin{subfigure}[t]{.31\linewidth}
    \includegraphics[width=\linewidth]{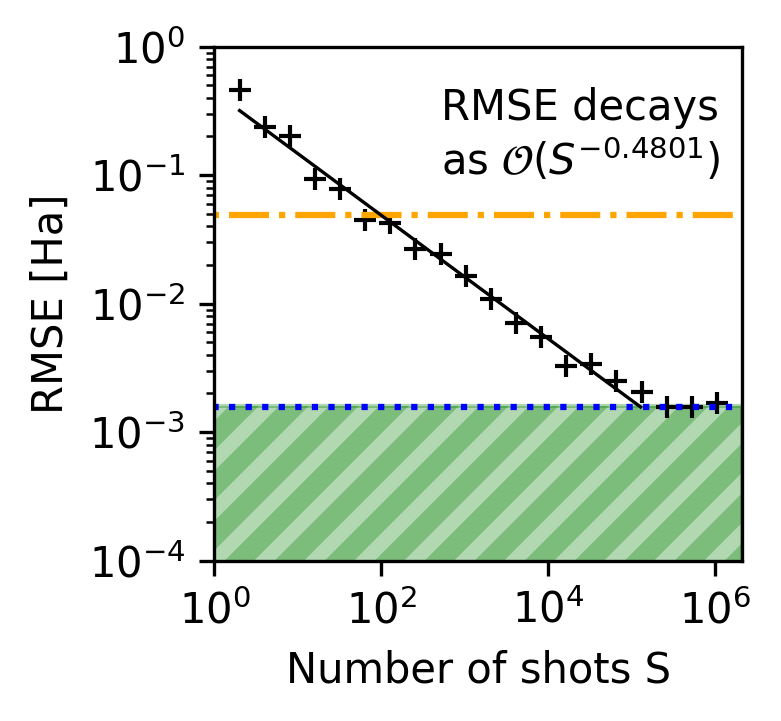}
    \caption{\ce{H2O} 7-qubit CS-VQE}\label{fig:H2O_7q_shot}
    \end{subfigure}
    \begin{subfigure}[t]{.31\linewidth}
    \includegraphics[width=\linewidth]{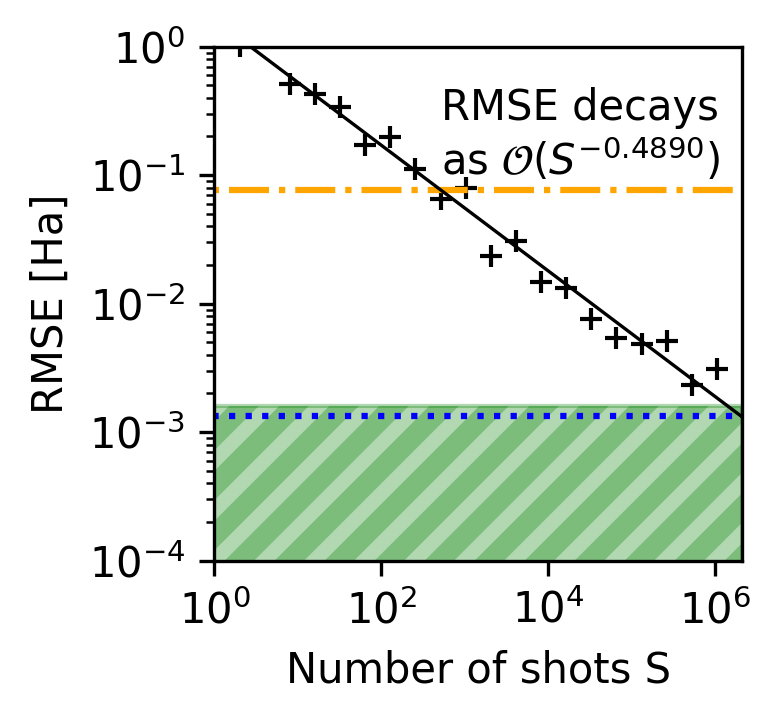}
    \caption{\ce{F2} 10-qubit CS-VQE}\label{fig:F2_10q_shot}
    \end{subfigure}
    \begin{subfigure}[t]{.31\linewidth}
    \includegraphics[width=\linewidth]{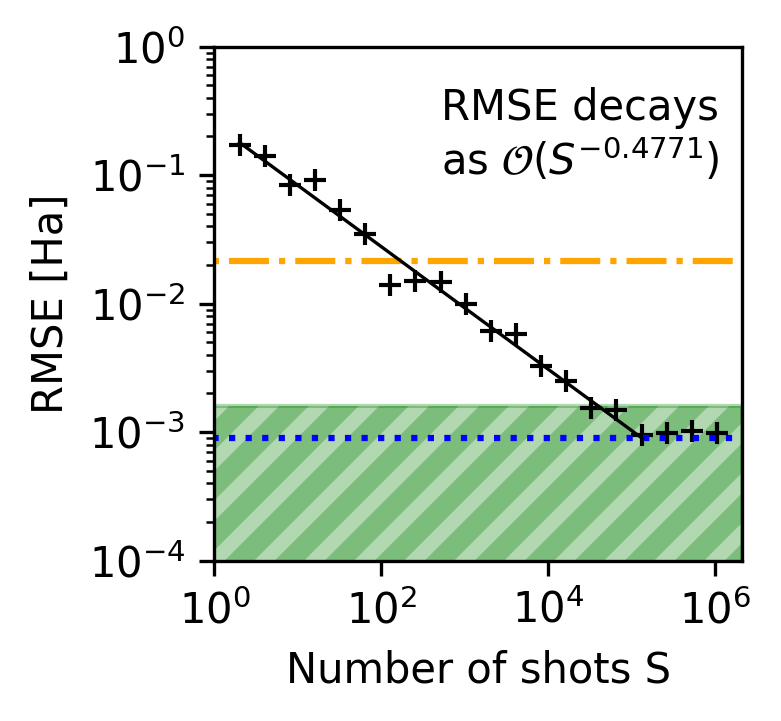}
    \caption{\ce{HCl} 4-qubit CS-VQE}\label{fig:HCl_4q_shot}
    \end{subfigure}
    \par\bigskip
    \begin{subfigure}[b]{0.9\linewidth}
    \includegraphics[width=\linewidth]{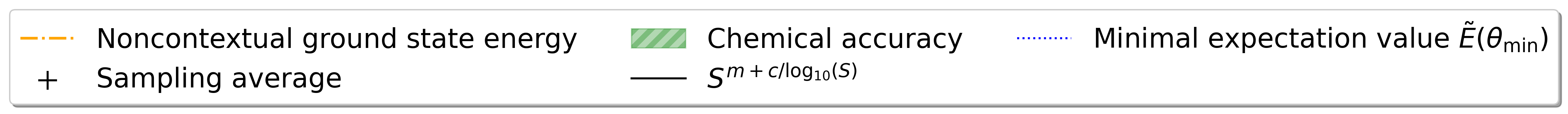}
    \end{subfigure}

\caption{Each of the plots \protect\ref{fig:Be_3q_shot} - \protect\ref{fig:HCl_4q_shot} correspond with \protect\ref{fig:Be_3q_conv} - \protect\ref{fig:HCl_4q_conv} above and illustrate the statistical effect of sampling noise at the optimal parametrization $\bm{\theta}_\mathrm{min}$ determined from the ADAPT-VQE statevector simulations in Figure \protect\ref{fig:QLM_conv_results}. We plot the root mean-square error (RMSE) for twenty `realizations' of the ground state energy estimate with $S \leq 10^6$ shots executed via IBM's QASM simulator; determining the line of best fit $m \cdot \log_{10}(S) + c$ with respect to the log-log data indicates a decay in error of $\mathcal{O}(S^m)$.}
\label{fig:QLM_shot_results}

\end{figure}

To achieve an absolute error of $\Delta > 0$, one should expect to perform $\mathcal{O}(\frac{1}{\Delta^2})$ shots (for each term of the Hamiltonian) \cite{peruzzo2014variational}. Conversely, suppose we are allocated a quantity $S \in \mathbb{N}$ of shots -- the obtained error should be of the order $\mathcal{O}(\frac{1}{\sqrt{S}})$. In order to increase estimate accuracy, we collected the Pauli terms into qubit-wise commuting (QWC) groups \cite{yen2020measuring} using the graph-colouring functionality of NetworkX \cite{osti_960616}; such groups may be measured simultaneously.

In Figure \protect\ref{fig:QLM_shot_results}, the number of shots $S = 2^n$ for $n = 0, \dots, 20$ carried out per QWC group is varied and we observe the root mean-square error (RMSE) over twenty realizations of the ground state energy estimate, plotted on a log-log scale. For clarity, note the \textbf{only} source of noise here is that which arises from statistical variation of the quantum circuit sampling -- we have \textbf{not} introduced hardware noise in the form of imperfect quantum gates or decoherence.

Two error regimes are observed, one of which is quite trivial: at high shot-counts we see a plateau resulting from the optimal error $| \Tilde{E}(\bm{\theta}_\mathrm{min}) - \epsilon_0 |$ being recovered. %It is interesting to note that in the presence of statistical noise, the VQE can cease to be variational! This is evidenced by the energy estimates occasionally 
To assess the convergence properties outside of this limiting region, we plot a line of best fit $m \cdot \log_{10} (S) + c$ among the data not exhibiting such behaviour; since the data is represented on a log-log scale, this corresponds with a decay in error of $\mathcal{O}(S^m)$. In each plot of Figure \protect\ref{fig:QLM_shot_results} we obtain $m \approx -0.5$, meaning the RMSE follows the predicted decay of $\mathcal{O}(\frac{1}{\sqrt{S}})$.

In every simulation bar \ce{F2}, chemical accuracy was achieved within $S=2^{20} \approx 10^6$ shots per QWC group. However, our shot budget could be reduced by implementing more advanced allocation strategies, for example according to the magnitude of Hamiltonian term coefficients \cite{arrasmith2020operator} or a classical shadow tomography approach \cite{huang2020predicting, hadfield2020measurements}.

\section{Conclusions}

We have placed CS-VQE on the theoretical footing of stabilizer subspace projections, which allows one to compare it against other qubit reduction techniques such as qubit tapering \cite{bravyi2017tapering, setia2020reducing}. Tapering defines a projection dependent on a symmetry of the full Hamiltonian and preserves the ground state energy exactly, whereas CS-VQE is approximate and projects onto a contextual subspace consistent with a symmetry of the noncontextual sub-Hamiltonian, augmented by an anticommuting contribution. In combination, the two techniques can effect a significant reduction in quantum resource requirements, as illustrated by Kirby et al. \cite{kirby2021contextual} and in Figure \protect\ref{fig:anz_depth}.

Previously, the only obstacle to building a CS-VQE framework that would be faithful to deployment on quantum devices was that of the ansatz, which has been addressed within this work. Furthermore, we demonstrated how CS-VQE may be combined with the ADAPT-VQE \cite{grimsley2019adaptive, tang2021qubit, shkolnikov2021avoiding, fedorov2021unitary} ansatz construction framework by applying our noncontextual projection to the operator pool; validation was presented in Figure \protect\ref{fig:QLM_conv_results} in which we achieved chemical accuracy for the suite of small molecules outlined in Table  \protect\ref{mol_suite}. This combination provides considerable flexibility in both qubit count and circuit depth, allowing one to identify a reduced problem that may be simulated on the available quantum resource.

A number of research questions concerning the scalability of CS-VQE remain; we recapitulate these here. Firstly, the success of CS-VQE is sensitive to the generator subset $\mathcal{F}$ one chooses to constrain in the stabilizer subspace projection. To date, the most effective method for choosing this subset has been a greedy-search heuristic necessitating $\mathcal{O}(N^{d+1})$ VQE simulations where $d \leq N$ is the search depth; this is expensive for NISQ hardware and there is room for more targeted heuristics. For example, we may draw on chemical intuition to inform the selection of a contextual subspace that captures information about the underlying electronic structure problem. The second obstacle lies in the approach taken to construct the noncontextual sub-Hamiltonian. There is currently no intuition as to what constitutes an effective choice here, although it should be noted the `optimal' noncontextual subset will not necessarily be that which minimizes the noncontextual ground state energy; some consideration of the resulting contextual subspaces must come into the construction of the noncontextual problem. We leave these issues for future work.

The natural next step is to execute this method on a NISQ computer, challenging the current best-in-class electronic structure simulations from Google, IonQ and IBM \cite{arute2020hartree, nam2020ground, eddins2021doubling}. To achieve this goal, CS-VQE could be combined with techniques of measurement reduction \cite{babbush2018low, wang2019accelerated, verteletskyi2020measurement, jena2019pauli, gokhale2019minimizing, yen2020measuring, huggins2021efficient, gokhale2020n, torlai2020precise, crawford2021efficient, huang2020predicting, hadfield2020measurements, izmaylov2019unitary, zhao2020measurement, bonet2020nearly, ralli2021implementation} and error mitigation \cite{temme2017error, li2017efficient, endo2018practical, kandala2019error, giurgica2020digital, he2020zero, endo2021hybrid, huggins2021virtual}.

Finally, we have written an open-source Python package that facilitates the stabilizer subspace projection techniques of this paper, with in-built tapering and CS-VQE functionality. We welcome the reader to make use of our code \cite{symmer2022}, which is freely available on GitHub.

\begin{acknowledgement}
T.W. and A.R. acknowledge  support from the Engineering and Physical Sciences Research Council (EP/S021582/1 and EP/L015242/1, respectively). T.W. also acknowledges support from CBKSciCon Ltd., Atos, Intel and Zapata. W.K. and P.J.L. acknowledge  support  by the NSF STAQ project (PHY-1818914). W. K. acknowledges support from the National Science Foundation, Grant No. DGE-1842474. P.V.C. is grateful for funding from the European Commission for VECMA (800925) and EPSRC for SEAVEA (EP/W007711/1). We would like to thank both Atos and the Leibniz Supercomputing Centre (LRZ), who each provided access to separate Atos Quantum Learning Machine (QLM) simulators that aided with the computational workload.
\end{acknowledgement}

\begin{suppinfo}
In the interest of reproducibility, we supply the tapering parameters, CS-VQE model data and noncontextual projection ans\"atze which permit chemical accuracy for the fewest number of qubits with respect to the molecular systems listed in Table \protect\ref{mol_suite}; the raw data for these results are supplied in ancillary files hosted on arXiv. This should provide sufficient information for the reader to reproduce Figures \protect\ref{fig:anz_depth} \protect\ref{fig:QLM_conv_results} and \protect\ref{fig:QLM_shot_results}.
\end{suppinfo}

\bibliography{main}
%\printbibliography

\end{document}

\newpage
\section{Circuit implementation}

Here we introduce the key concepts that are required to implement our CS-VQE ansatz as a quantum circuit, withholding a thorough introduction to the topic (the reader is referred to Nielsen and Chuang for this \cite{nielsen2010quantum}. 

\subsection{Trotterization}\label{trotter}

For operators $A, B \in \mathscr{B}(\mathscr{H})$, we have $e^{A+B} = e^A e^B$ if and only if $A$ commutes with $B$ (i.e. $[A, B]=0$), contrary to the familiar rules of exponentiation for numbers; in the more general case, the Lie product formula states
\begin{equation}
    e^{A+B} = \lim_{n\rightarrow\infty} \Big(e^{A/n} e^{B/n}\Big)^n.
\end{equation}
The technique of \textit{Suzuki-Trotter} truncates the above limit at some $n_\mathrm{T} \in \mathbb{N}$ (the \textit{Trotter number}) to yield an approximation of the exponentiated operator sum. Given a sum of operators $\sum_{k} \theta_k A_k, \theta_k \in \mathbb{R}$, we may write the first-order Trotter expansion
\begin{equation}\label{trotterexpand}
    e^{i \sum_{k} \theta_k A_k} \approx \bigg( \prod_k e^{i \frac{\theta_k}{n_\mathrm{T}} A_k} \bigg)^{n_\mathrm{T}},
\end{equation}
where the exponentiated Pauli operator $e^{i \frac{\theta_k}{n_\mathrm{T}} A_k}$ may be performed in-circuit as described in \ref{sec_exp_P} using $\mathcal{O}(N)$ native quantum gates.

Used in combination with VQE, $n_\mathrm{T} = 1$ is often sufficient to achieve high levels of precision since we expect the optimizer to produce different coefficients that counteract the Trotter error. It has also been observed that the ordering of Trotter terms has an impact on errors \cite{tranter2019ordering} -- this will not be explored here, however it is a possible consideration for future research. There is contention over Trotterized UCC and whether it is well-defined as an ansatz, since the ordering of terms can induce variations in error beyond chemical accuracy, particularly for strongly correlated systems \cite{grimsley2019trotterized}. %In this regard, it is argued that ans\"atze whose ordering is deterministic, such as ADAPT-VQE, are preferable.

\subsection{Exponentiating Pauli strings}\label{sec_exp_P}

Given a Pauli operator $P \in \mathcal{P}_N$ and some angle $\theta \in \mathbb{R}$, we would like to implement the exponential $e^{i \theta P}$ as a quantum circuit; this can be achieved with $\mathcal{O}(N)$ gates. We shall first assume that $P$ consists of just Pauli $I, Z$ operators, with the qubit positions acted upon by $Z$ indexed with the set $\mathcal{I}_Z$. Observe that
\begin{equation}\label{expZ}
\begin{aligned}
    e^{i \theta Z^{\otimes \mathcal{I}_Z}} \ket{\bm{b}} 
    ={} & \Big(\cos\theta + i (-1)^{\rho_{\mathcal{I}_Z}(\bm{b})} \sin\theta\Big) \ket{\bm{b}}\\ 
    ={} & e^{i \theta (-1) ^{\rho_{\mathcal{I}_Z}(\bm{b})}} \ket{\bm{b}}.
\end{aligned}
\end{equation}
where we have omitted the qubit positions that are identity. Therefore, we may realise this operation by storing $\rho_{\mathcal{I}_Z}(\bm{b})$, the parity of $\ket{\bm{b}}$ over $\mathcal{I}_Z$, in one of the qubits and applying to it an $R_Z$ gate, defined by
\begin{equation}
    R_Z (\phi) = \begin{pmatrix} e^{i\phi/2} & 0 \\ 0 & e^{-i\phi/2} \end{pmatrix}.
\end{equation}
In \eqref{expZ}, even parity results in a phase $e^{i\theta}$ whereas odd parity yields $e^{-i\theta}$ -- these are obtained by application of $R_Z(2\theta)$ to each of $\ket{0}, \ket{1}$, respectively. The parity computation is accomplished via a `cascade' of CNOT gates between adjacent qubits. We are now in a position to explicitly write down a quantum circuit that effects the exponentiation of a Pauli string consisting of $I, Z$ operators, presented in Figure \protect\ref{fig:expZ_circuit}.

\begin{figure}[b]
    \centering
    \input{exp_Z}
    \caption{Circuit to realise the exponential $e^{i \theta P}$ where $P$ consists of just Pauli $I, Z$ operators. The qubits set to identity are omitted from the diagram.}
    \label{fig:expZ_circuit}
\end{figure}

This is the basic building block for exponentiating an arbitrary Pauli string, as we are able to make a change of basis from Pauli $X, Y$ operators to a Pauli $Z$. Note that $X = H Z H$ and $ Y = S H Z H S^\dag$ where $S = R_Z(\pi/2)$. Once this transformation has been applied, the problem is reduced to that of before and we perform the same parity computation and $Z$ rotation. We end with the reverse transformation taking us back into the original basis -- the complete circuit is presented in Figure \protect\ref{fig:expP_circuit}. The total number of gates is at most $6N-1$, achieved only when $P$ consists solely of Pauli $Y$ operators acting upon all of the $N$ qubit positions.

\begin{figure}
    \centering
    \input{exp_P}
    \caption{Circuit to realise the exponential $e^{i \theta P}$ where $P$ is an arbitrary Pauli string. The qubits set to identity are omitted from the diagram.}
    \label{fig:expP_circuit}
\end{figure}

It is noted that in the Trotterized circuit there will be many blocks of exponentiated Pauli strings of the form displayed in Figure \protect\ref{fig:expP_circuit}, and it is possible that gate cancellation will occur between adjacent blocks where we have $SS^\dag = H^2 = \mathbbm{1}$. Furthermore, the implementation presented here is not unique -- different circuits yielding the same quantum states are possible.

\section{Noncontextual objective function}\label{nc_obj_func}

The CS-VQE method relies on solving a classical objective function that defines the energy spectrum of the noncontextual Hamiltonian
\begin{equation}
    H_{\mathcal{T}_\mathrm{nc}} = \sum_{P \in \mathcal{T}_\mathrm{nc}} h_{P} P.
\end{equation}
The objective function is constructed from the decomposition 
\begin{equation}
    \mathcal{T}_\mathrm{nc} = \mathcal{S} \cup \mathcal{C}_1 \cup \dots \cup \mathcal{C}_M,
\end{equation}
in which one observes that the sum over $\mathcal{T}_\mathrm{nc}$ may be separated into individual summations over the symmetry terms $\mathcal{S}$ and the anticommuting classes $\mathcal{C}_i$ for $i \in \{1, \dots, M\}$, i.e. $\sum_{P \in \mathcal{T}_\mathrm{nc}} = \sum_{P \in \mathcal{S}} + \sum_{i=1}^M \sum_{P \in \mathcal{C}_i}$. Furthermore, since $\mathcal{G}$ generates the symmetry group $\mathcal{S}^\prime$ and for any $C_i^\prime \in \mathcal{C}_i$ we have $C_i C_i^\prime \in \mathcal{S}^\prime$ by construction, it holds that $\mathcal{S} \subset \overline{\mathcal{G}}$ and $\mathcal{C}_i \subset C_i\overline{\mathcal{G}} \coloneqq \{C_i P | P \in \overline{\mathcal{G}}\}$. This means we may instead sum over the completion of $\mathcal{G}$, noting that any summand not appearing in $\mathcal{T}_\mathrm{nc}$ will have zero coefficient:
\begin{equation}
    H_{\mathcal{T}_\mathrm{nc}} = \sum_{P \in \overline{\mathcal{G}}} \bigg(h_{P}^\prime + \sum_{i=1}^M h_{P,i} C_i \bigg) P,
\end{equation}
where
\begin{equation}
    h_{P}^\prime = 
    \begin{cases} 
        h_{P}, & P \in \mathcal{S} \\
        0, & \mathrm{otherwise}
    \end{cases},\;
    h_{P, i} = 
    \begin{cases} 
        h_{C_i P}, & C_i P \in \mathcal{C}_i \\
        0, & \mathrm{otherwise}
    \end{cases}.
\end{equation}

We now wish to use this reformulation of the noncontextual Hamiltonian to evaluate expectation values in terms of the parameters $\bm{\nu} \in \{\pm1\}^{\times |\mathcal{G}|}$, denoting eigenvalue assignments to the generators $\mathcal{G}$ (sectors), and $\bm{r} \in \mathbb{R}^M$, a unit vector weighting the class representatives. We define expectation values with respect to a joint probability distribution
\begin{equation}\label{nc_prob_dist}
    P(\bm{\nu}^\prime, \bm{r}^\prime | \bm{\nu}, \bm{r}) \coloneqq \frac{\delta_{\bm{\nu}, \bm{\nu}^\prime}}{2^M} \prod_{i=1}^M |r_i + r_i^\prime|,
\end{equation}
which is a special-case of the phase-space distribution given by Spekkens \cite{Spekkens2016}. Kirby \& Love \cite{kirby2020classical} proved the generators take expectation values 
\begin{equation}
    \braket{G}_{(\bm{\nu}, \bm{r})} = \nu_{f(G)} \;\forall\, G \in \mathcal{G},
\end{equation}
recalling that $f: \mathcal{G} \to \mathcal{I}_\mathrm{stab}$ is a mapping onto distinct qubit indices, and class representatives
\begin{equation}
    \braket{C_i}_{(\bm{\nu}, \bm{r})} = r_i \;\forall\, i \in \{1, \dots, M\}.
\end{equation}
Observe that 
\begin{equation}
    \braket{C(\bm{r})}_{(\bm{\nu}, \bm{r})} = \sum_{i=1}^M r_i^2 = |\bm{r}| = +1,    
\end{equation}
implying any quantum state consistent with the noncontextual state $(\bm{\nu}, \bm{r})$ must be stabilized by $C(\bm{r})$.

Putting everything together, we can express the expectation value of our noncontextual Hamiltonian as
\begin{equation}\label{classical_objective}
\begin{aligned}
    \braket{H_{\mathcal{T}_\mathrm{nc}}}_{(\bm{\nu}, \bm{r})} 
    ={} & \sum_{P \in \overline{\mathcal{G}}} \bigg(h_{P}^\prime + \sum_{i=1}^M h_{P,i} \braket{C_i}_{(\bm{\nu}, \bm{r})} \bigg) \braket{P}_{(\bm{\nu}, \bm{r})} \\
    ={} & \sum_{P \in \overline{\mathcal{G}}} \bigg(h_{P}^\prime + \sum_{i=1}^M h_{P,i} r_i \bigg) \prod_{G \in \mathcal{G}_{P}} \nu_{f(G)},
\end{aligned}
\end{equation}
where $\mathcal{G}_{P} \subset \mathcal{G}$ satisfies $P = \prod_{G \in \mathcal{G}_{P}} G$. Note also that we have used the fact  $\braket{AB} = \braket{A}\braket{B}$ when $[A, B] = 0$.

Taken as a classical optimization problem, minimizing the objective function
\begin{equation}\label{eta_func}
    \eta(\bm{\nu}, \bm{r}) \coloneqq \braket{H_{\mathcal{T}_\mathrm{nc}}}_{(\bm{\nu}, \bm{r})} 
\end{equation}
is NP-complete in general. Despite this, we expect typical instances to be heuristically solvable by classical methods \cite{kirby2020classical}. If one fixes the $\pm1$ eigenvalue assignments $\bm{\nu}$ -- a case of identifying the correct symmetry sector -- this becomes a convex optimization problem over points of the unit $(M-1)$-sphere.

\section{ADAPT-VQE}

Adaptive Derivative-Assembled Pseudo-Trotter (ADAPT) VQE \cite{grimsley2019adaptive} is a contemporary method of ansatz construction that retains some problem specificity while resulting in dramatically reduced circuit depths compared to conventional chemically-motivated approaches such as Unitary Coupled Cluster \cite{taube2006new, romero2018strategies}. The implementation we describe here is of the qubit-ADAPT-VQE \cite{tang2021qubit} variant, which is centred around an operator pool $\mathcal{O} \subset \mathcal{P}_N$; this could for example consist of excitation terms obtained from a Jordan-Wigner encoded coupled-cluster calculation. Terms are selected from the operator pool to append to a dynamically expanding ansatz in line with a gradient-based argument which we describe in Section \ref{ADAPTalg}. 

\subsection{Calculating gradients}

Take an operator $A(\bm{\theta}) = \sum_{k} \theta_k P_k$ for $\theta_k \in \mathbb{R}, P_k \in \mathcal{P}_N$, a reference state $\ket{\psi_\mathrm{ref}} \in (\mathbb{C}^2)^{\otimes N}$ and define the ansatz state 
\begin{equation}
    \ket{\psi (\bm{\theta})} = e^{i A(\bm{\theta})} \ket{\psi_\mathrm{ref}}.
\end{equation}
Application of the chain rule yields
\begin{equation}
    \frac{\partial}{\partial\theta_k} \ket{\psi(\bm{\theta})} = i P_k \ket{\psi(\bm{\theta})}
\end{equation}
and combining this with the product rule, we have
\begin{equation}
\begin{aligned}
    \frac{\partial}{\partial\theta_k} \braket{H}_{\psi(\bm{\theta})} 
    ={} & \bra{\psi(\bm{\theta})} [-i P_k H + i H P_k ] \ket{\psi(\bm{\theta})} \\
    ={} & i \braket{[H, P_k]}_{\psi(\bm{\theta})};
\end{aligned}
\end{equation}
this is the formulation proposed by Grimsley et al. \cite{grimsley2019adaptive}. However, if one observes that
\begin{equation}
\begin{aligned}
    i [H, P_k] 
    ={} & \frac{1}{2} \big[(I-iP_k)H(I+iP_k) - (I+iP_k)H(I-iP_k)\big]\\
    ={} & e^{- i \frac{\pi}{4} P_k} H e^{i \frac{\pi}{4} P_k} - e^{i \frac{\pi}{4} P_k} H e^{- i \frac{\pi}{4} P_k}
\end{aligned}
\end{equation}
then we recover the parameter shift rule of Parrish et al. \cite{parrish2019hybrid}, namely
\begin{equation}\label{paramshift}
    \frac{\partial}{\partial\theta_k} \braket{H}_{\psi(\bm{\theta})} = \braket{H}_{e^{i \frac{\pi}{4} P_k}\psi(\bm{\theta})} - \braket{H}_{e^{-i \frac{\pi}{4} P_k} \psi(\bm{\theta})}.
\end{equation}
Therefore, we may evaluate the expectation value $\braket{H}_{\psi(\bm{\theta})}$ at $\theta_k + \frac{\pi}{4}$ and $\theta_k - \frac{\pi}{4}$, with the partial gradient with respect to $\theta_k$ being their difference. We opt for the latter method of gradient calculation; this allows us to avoid storing a large collection of commutators in memory -- each requiring decomposition into qubit-wise commuting measurement groups -- at the expense of one additional expectation value calculation per partial gradient.

\subsection{The algorithm}\label{ADAPTalg}

We shall define the algorithm iteratively, initializing a reference state $\ket{\psi_0} \coloneqq \ket{\psi_\mathrm{ref}}$ and describing mathematically how one obtains $\ket{\psi_{k+1}}$ from the $k$-th ADAPT-VQE cycle. Let us define the function 
\begin{equation}
    g^{(k)}_{P}(\theta) \coloneqq \bra{\psi_k} e^{-i \theta P}  H e^{i \theta P} \ket{\psi_k}
\end{equation}
and
\begin{equation}
\begin{aligned}
    G_P^{(k)} (\theta) 
    \coloneqq{} & \frac{\partial}{\partial \theta} \braket{H}_{e^{i \theta P} \psi_k}  \\
    ={} & g^{(k)}_{P}\Big(\theta + \frac{\pi}{4}\Big) - g^{(k)}_{P}\Big(\theta - \frac{\pi}{4}\Big)
\end{aligned}
\end{equation}
by \eqref{paramshift}.

Then, identifying the pool operator that yields the largest gradient in magnitude at $\theta = 0$,
\begin{equation}
   P_{k+1} = \argmax_{P \in \mathcal{O}} \Big| G_P^{(k)} (0)   \Big|,
\end{equation}
we construct the expanded ansatz 
\begin{equation}
    \ket{\psi_{k+1} (\bm{\theta}^{(k+1)})} = e^{i \theta_{k+1} P_{k+1}} \ket{\psi_{k} (\bm{\theta}^{(k)})}.    
\end{equation}
We now perform a VQE simulation to find
\begin{equation}
    \bm{\theta}^{(k+1)}_\mathrm{min} = \argmin_{\bm{\theta}^{(k+1)} \in \mathbb{R}^{k+1}} \braket{H}_{\psi_{k+1} (\bm{\theta}^{(k+1)})}
\end{equation}
and finally set
\begin{equation}
    \ket{\psi_{k+1}} =  \ket{\psi_{k+1}(\bm{\theta}^{(k+1)}_\mathrm{min})}.
\end{equation}
This process is iterated until some termination condition is satisfied, for example when the largest gradient $\max_{P \in \mathcal{O}} \big| G_P^{(k)} (0) \big|$ falls below a predefined threshold.

We note a few properties of ADAPT-VQE. First of all, in theory, the minimized energy $\braket{H}_{\psi_{k} (\bm{\theta}^{(k)}_{\mathrm{min}})}$ will decrease monotonically with increasing $k \in \mathbb{N}$. In practice, the ansatz complexity also increases and therefore optimization becomes increasingly demanding, meaning it is possible the true minimum is not found. In such cases, it is possible for the $(k+1)$-th iterate to yield higher energy than the $k$-th. 

Secondly, the operator pool $\mathcal{O}$ is never exhausted, meaning it is possible to select the same term more than once. This gives rise to the `Pseudo-Trotter' description of ADAPT-VQE, in the sense that it somewhat resembles the Trotter expansion \eqref{trotterexpand} but with only select terms duplicated.

\newpage
\bibliography{supplement}